\documentclass[a4paper,fleqn]{cas-sc}

\usepackage[numbers]{natbib}
\usepackage{caption}
\usepackage{ulem}
\usepackage[utf8]{inputenc}

\def\tsc#1{\csdef{#1}{\textsc{\lowercase{#1}}\xspace}}
\tsc{WGM}
\tsc{QE}
\begin{document}
\let\WriteBookmarks\relax
\def\floatpagepagefraction{1}
\def\textpagefraction{.001}

\shorttitle{}    

\shortauthors{}  

\title [mode = title]{Magnetization, Susceptibility, and Entanglement Entropy in Antiferromagnetic Star-Shaped Homometallic Tetranuclear and Pentanuclear Cu(II) Complexes}  



 \author[1]{G. Mirzoyan}[%type=editor,
       orcid=0009-0002-8565-2456,
%       gplus=<gplus id>
]

\author[2,3]{Vl. Papoyan}[orcid=0000-0001-6854-4723]

%\cormark[1]


%\ead{}



\affiliation[1]{organization={Center for the Advancement of Natural Discoveries using Light Emission (CANDLE)},
            addressline={}, 
            city={Yerevan},
          citysep={}, % Uncomment if no comma needed between city and postcode
            postcode={0022}, 
            %state={},
            country={Armenia}}

\author[4]{A. Trombettoni}%[]

\author[4]{N. Titov}[orcid=0009-0004-0259-4655]
\cortext[1]{Corresponding author}
\ead{nikita.titov@units.it}

\author[5]{L. Ananikyan}

\author[1,5]{N. Ananikian}

% Footnote of the second author

% Email id of the second author
%\ead{}



\affiliation[2]{organization={Joint Institute for Nuclear Research (JINR), Bogoliubov Laboratory of Theoretical Physics},
            %addressline={}, 
            city={Dubna},
          citysep={}, % Uncomment if no comma needed between city and postcode
            postcode={141980}, 
            %state={},
            country={Russia}}

\affiliation[3]{organization={Dubna State University},
            %addressline={}, 
            city={Dubna},
          citysep={}, % Uncomment if no comma needed between city and postcode
            postcode={141982}, 
            %state={},
            country={Russia}}
            
\affiliation[4]{organization={Department of Physics, University of Trieste},
            %addressline={}, 
            city={Trieste},
          citysep={}, % Uncomment if no comma needed between city and postcode
            postcode={34127}, 
            %state={},
            country={Italy}}

\affiliation[5]{organization={A.I. Alikhanyan National Science Laboratory , Theoretical Physics Division},
            %addressline={}, 
            city={Yerevan},
          citysep={}, % Uncomment if no comma needed between city and postcode
            postcode={0036}, 
            %state={},
            country={Armenia}}

%\cortext[1]{Corresponding author}



\begin{abstract}
We present a theoretical investigation of the magnetization, magnetic susceptibility, and quantum entanglement entropy in antiferromagnetic star-shaped tetranuclear ($Cu_4$) and pentanuclear ($Cu_5$) Cu(II) clusters, motivated by experimentally realized molecular spin systems. In both cases, the study presents a discussion of the comparison between experimental data and theoretical analytical calculations of magnetic susceptibility as a function of temperature.

The correlation between magnetization jumps and magnetic susceptibility peaks, along with the presence of entanglement entropy plateaus as a function of external magnetic fields, is analyzed in the low-temperature regime.

Overall, the results demonstrate  that the cluster size, geometry, and interaction anisotropy play a crucial role in shaping both the thermodynamic and the entanglement properties. In addition, the study is based on antiferromagnetic Heisenberg spin Hamiltonians, with the inclusion of the Dzyaloshinskii–Moriya interaction.
\end{abstract}


%\item 
%\item 
%\item 

% Keywords
\begin{keywords}
 Tetranuclear Cu(II) complexes\sep Pentanuclear Cu(II) complexes\sep Magnetization\sep Susceptibility\sep Entropic Entanglement\sep Quantum Correlations
\end{keywords}

\maketitle

\section{Introduction}

Studying the interplay between geometric frustration, exchange interactions, and anisotropic couplings provides guidance for designing molecules with tailored magnetic properties\cite{Kahn1993,Diep2013,Lacroix2011,Sachdev1999}.  Moreover, understanding quantum entanglement in molecular spin clusters is not only of theoretical interest but also of practical relevance for quantum technologies. The 1/3 magnetization plateau has been experimentally observed in natural antiferromagnetic azurite compounds \cite{Kikuchi2005_PRL}. Moreover, both experimental and theoretical investigations have been conducted on natural antiferromagnetic azurite compounds \cite{Gu2007,Takano1996,Tonegawa2001,Rule2008,Kang2009,Jeschke2011,Ananikian2012,Okamoto1999,Rojas2012,ArianZad2017,ArianZad2018,ArianZad2019a}. The magnetization properties and correlation behavior of complexes are investigated in both classical and quantum cases, with or without the Dzyaloshinsky-Moriya interaction in high-spin models \cite{Ananikian1991Phase,Hovhannisyan2009Partition,Izmailian1994General,Ananikian2023Quantum,Hovhannisyan2016Partition,
ArianZad2025Quantum}.
 
 Low-dimensional quantum spin models provide an ideal framework for studying non-classical correlations, quantum phase transitions, and thermodynamic properties in strongly correlated systems.
 In translationally invariant systems, the Fourier transform allows one to decouple momentum modes, but in the absence of translational invariance the theoretical analysis is more challenging \cite{Monceau2011,Titov2026on,Titov2026laplacian}. However, by considering physical models defined on non-translationally invariant structures one is able to study new phenomena not present in their lattice counterparts \cite{Brunelli2004,Tsvelik2013,Crampe2013}. These phenomena may stem from, e.g., the non-trivial geometry of the graph, localization effects, or modification of the density of states. A plethora of physical models defined on graphs have been studied in a variety of contexts \cite{Buonsante2002,Rammal1983,Burioni2005,Millan2021,Lucci2022,Burioni2000}. Among graphs, the star graph, in which several lines, called  "legs", merge at a point (or in a region), provides a simple yet paradigmatic example of the effects of the breakdown of translational invariance, which affects local properties around the center (or junction) of the star \cite{Brunelli2004,Buonsante2002,Lucci2022,Burioni2000}. The spectrum of the adjacency matrix \cite{Harary1969} of a star graph, i.e., of the matrix $A_{ij}$ having values $1$ if $i$ and $j$ are nearest neighbors and $0$ otherwise, has a localized ground state, giving rise to the spatial condensation of ideal bosons when they are defined on the star \cite{Brunelli2004,Burioni2000}. Star graphs or multiple connected star graphs have been studied in the context of superconducting junctions \cite{Silvestrini2007,Lorenzo2014,Romeo2022}. Clearly, the shorter the legs of the star, the more pronounced the effects of the presence of the center. Given the presence of studies of ultracold atoms and superconducting devices on star graphs, it would be desirable to have solid-state magnetic systems or molecules in which one can study the effects of the structure on quantum properties.
 
 Quantum entanglement in low-dimensional spin systems has attracted considerable attention due to its fundamental role in quantum information science and condensed matter physics. In a number of material systems, including copper-based complexes, it has been demonstrated that entanglement measures such as concurrence and entanglement entropy are very useful theoretical tools to characterize their quantum properties. The study of quantum entanglement in inhomogeneous structures provides valuable insight into how spatial structure and connectivity influence quantum correlations in these systems \cite{Franca2008,Calabrese2012,Bentsen2019,Buruage2026}. In particular, it is known that star graphs can host highly entangled states, due to the large connectivity of the center site which acts as an impurity connecting different legs \cite{Tsvelik2013,Giuliano2008,Tsvelik2014,Giuliano2025}.  
Thermal non-classical correlations quantified by concurrence entanglement, quantum coherence, and teleportation fidelity were %exactly 
examined for ferromagnetic and mixed ferromagnetic-antiferromagnetic states in the square four-qubit Heisenberg $XXZ$ $Cu_{4}^{II}$ complex \cite{Benabdallah2022}. The magnetic properties of a number of dinuclear and trinuclear antiferromagnetic copper (II) complexes with various bridging ligands were studied. The research applied theoretical models of antiferromagnetic properties and experimental data to analyze magnetization plateaus, magnetic susceptibility, and entanglement entropy \cite{Rojas2025}. It has been found that entanglement measures based on spin correlation functions and magnetic susceptibility can serve as indicators of entanglement for a wide class of systems \cite{Glaser2003, Wiesniak2005}. The structures and magnetic properties of antiferromagnetic star-shaped homometallic tetranuclear Cu(II) clusters of the composition $[[Cu^{II}L^{1}]_2[Cu^{II}L^{1}Cu^{II}(H_{2}O)_{3}]](ClO_{4})_{2}$ (\textbf{1}) derived from the hexadentate Schiff base compartmental ligand $N$,$N'$-ethylenebis(3-ethoxysalicylaldimine) ($H_{2}L^{1}$) have been described \cite{Nayak2006}. A similar structure and magnetic properties of the antiferromagnetic star-shaped homometallic pentanuclear $Cu^{II}$ cluster were obtained using a semi-flexible aminotriazine-based bis-methylpyridine ligand \cite{Chiang2016}.

In this work, we investigate the magnetization, magnetic susceptibility, and entanglement entropy in star-shaped antiferromagnetic tetranuclear $Cu_4$ and pentanuclear $Cu_5$ clusters modeled after experimentally synthesized compounds with and without the Dzyaloshinskii-Moriya (DM) interaction. Throughout this work we restrict the analysis to the low-temperature regime, where magnetic exchange and Dzyaloshinskii–Moriya interactions dominate the thermodynamic and entanglement properties of the clusters. At higher temperatures, spin-phonon coupling is expected to become increasingly relevant, modifying both transport and quantum-correlation properties, as demonstrated for anisotropic XY-type spin models \cite{Lima2020,Lima2021}. Entanglement properties of  tetranuclear $Cu_4$ clusters in the absence of the Dzyaloshinskii–Moriya interaction were previously analyzed by Karl\'ová and Strečka \cite{Karlova2023} using the negativity as an entanglement measure. We extend these results through the inclusion of the Dzyaloshinskii–Moriya interaction and the comparison with pentanuclear $Cu_5$ clusters.

The article is structured as follows. In Section 2, we present the model Hamiltonian and the theoretical framework. Section 3 is devoted to the analysis of the magnetic susceptibility of the clusters, primarily aimed at a comparison with the available experimental results. The field dependence of entanglement entropy, magnetization, and magnetic susceptibility is investigated, with particular emphasis on the emergence of plateau structures and the identification of their characteristic peaks, as discussed in Section 4. Finally, Section 5 summarizes the main results and conclusions.

\section{Models}

\begin{figure}[]
\centering
\includegraphics[width=0.48\textwidth]{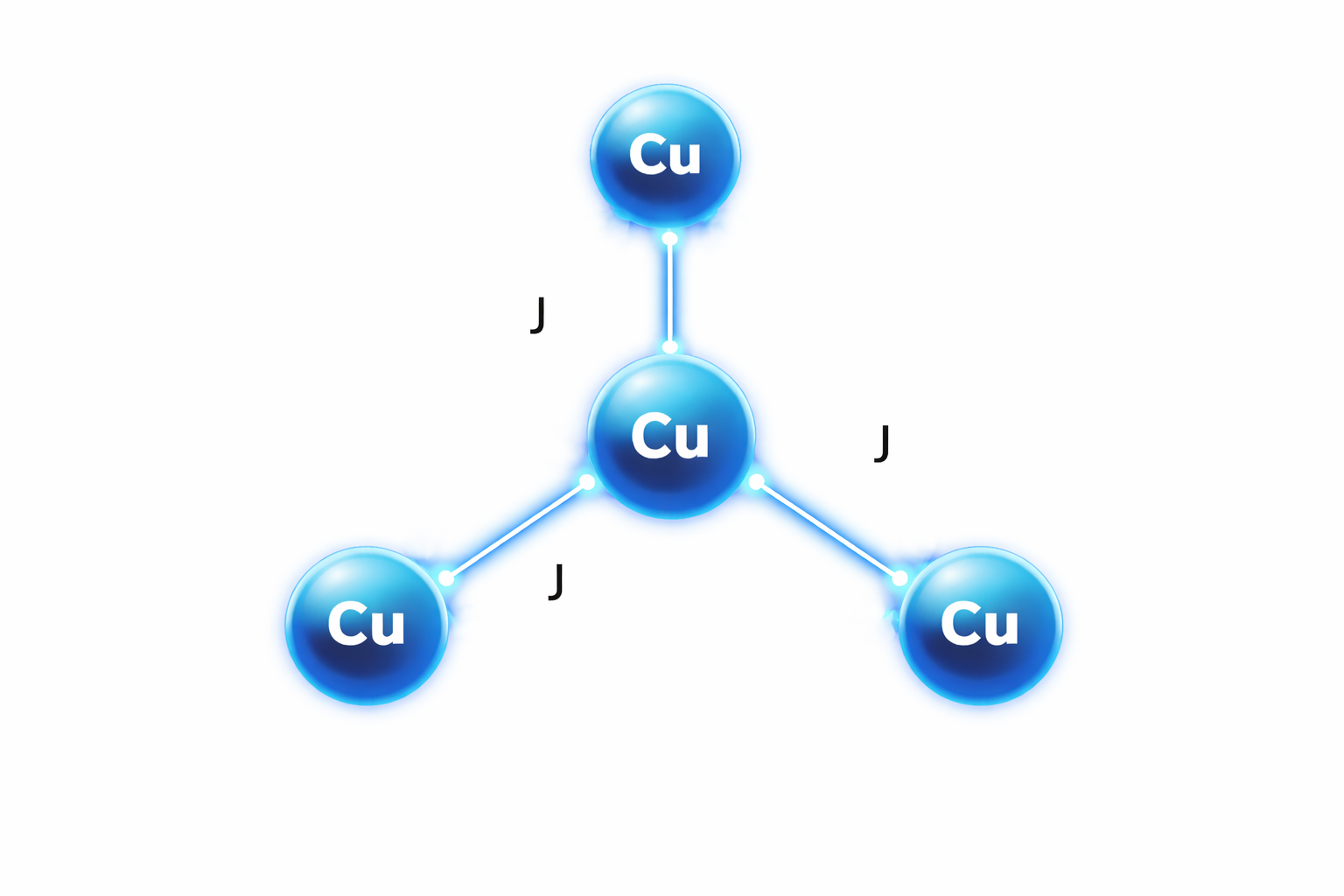}
\hfill
\includegraphics[width=0.48\textwidth]{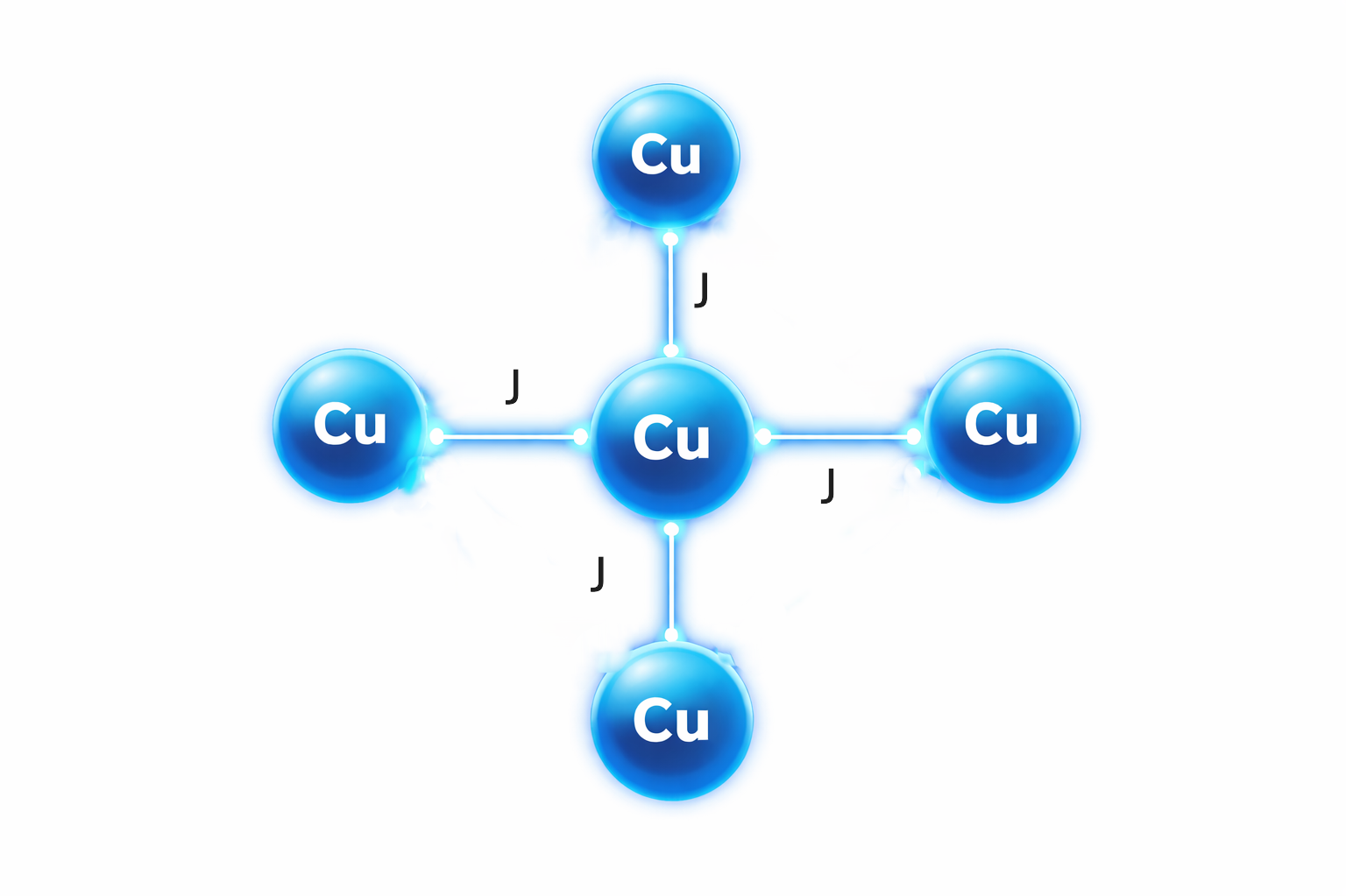}
\caption{Schematic representation of the copper clusters: (left) the tetranuclear $Cu_4$ cluster, where the central Cu(II) ion interacts antiferromagnetically with three peripheral Cu(II) ions; (right) the pentanuclear $Cu_5$ cluster, where the central Cu(II) ion mediates antiferromagnetic interactions with four peripheral Cu(II) ions.}
\label{fig:Cu_clusters}
\end{figure}

In this work, we consider two classes of star-shaped homometallic copper(II) clusters, inspired by experimental realizations \cite{Nayak2006, Chiang2016}. The first is a tetranuclear $Cu_4$ cluster, with the general formula $[[Cu^{II}L^{1}]_2[Cu^{II}L^{1}Cu^{II}(H_{2}O)_{3}]](ClO_{4})_{2}$, in which a central Cu(II) ion interacts with three peripheral Cu(II) ions arranged in a planar star geometry (Fig.~\ref{fig:Cu_clusters}). The second system is a pentanuclear $Cu_5$ cluster, [$Cu_5$(OH)$_4$(H$_2$dpmta)$_2$(ClO$_4$)$_4$(H$_2$O)$_4$](ClO$_4$)$_2 \cdot$ Et$_2$O$\cdot$5iPrOH, where four peripheral Cu(II) ions surround a central Cu(II) ion in a symmetric star-shaped arrangement (Fig.~\ref{fig:Cu_clusters}). Each Cu(II) ion carries spin $S=1/2$, and we assume isotropic antiferromagnetic exchange interactions between the central and peripheral ions, while direct interactions between the peripheral ions are neglected due to geometric constraints.

To enable a possible comparison with experimental measurements, we first consider the isotropic Heisenberg Hamiltonian in the presence of an external magnetic field $h$, neglecting anisotropic interactions.

For the $Cu_5$ cluster, the Hamiltonian is given by
\begin{equation}
H = -2J \sum_{i=2}^{N} \vec{S}_1 \cdot \vec{S}_i 
- g \mu_B h \sum_{i=1}^{N} S_i^z,
\label{eq:hamCu5}
\end{equation}
whereas for the $Cu_4$ cluster it takes the form
\begin{equation}
H = -J \sum_{i=2}^{N} \vec{S}_1 \cdot \vec{S}_i 
- g \mu_B h \sum_{i=1}^{N} S_i^z.
\label{eq:hamCu4}
\end{equation}
Here, $\vec{S}_1$ denotes the central spin, $\vec{S}_i$ correspond to the peripheral spins, $J<0$ is the antiferromagnetic exchange coupling constant, $g$ is the Land\'e factor and $\mu_B$ is the Bohr magneton. The total number of spins is $N=5$, $J = -82.5cm^{-1}$, $g = 2.16$ for the $Cu_5$ and $N=4$, $J = -17.4cm^{-1}$, $g = 2.087$ for the $Cu_4$.

The energy spectrum of these Hamiltonians can be obtained analytically. For the $Cu_5$ cluster, the two lowest-energy eigenstates are given by
\begin{equation}
E_{1} = -2J - \frac{5}{2}\mu_{B}gh, \quad |\psi_{1}\rangle = |\uparrow\uparrow\uparrow\uparrow\uparrow\rangle,
\end{equation}
\begin{equation}
E_{2} = 3J - \frac{3}{2}\mu_{B}gh,
\end{equation}
\begin{equation}
|\psi_{2}\rangle = \frac{1}{2\sqrt{5}}\left(4|\downarrow\uparrow\uparrow\uparrow\uparrow\rangle
-|\uparrow\downarrow\uparrow\uparrow\uparrow\rangle
-|\uparrow\uparrow\downarrow\uparrow\uparrow\rangle
-|\uparrow\uparrow\uparrow\downarrow\uparrow\rangle
-|\uparrow\uparrow\uparrow\uparrow\downarrow\rangle\right).
\end{equation}

The crossing of these energy levels defines a critical magnetic field
\begin{equation}
h_c =- \frac{5J}{\mu_B g} = 409.109~\text{T},
\end{equation}
which signals a quantum phase transition between states with different total magnetization.

For the $Cu_4$ cluster, the corresponding eigenvalues and eigenstates are
\begin{equation}
E_{1} = -\frac{3}{4}J - 2\mu_{B}gh, \quad |\psi_{1}\rangle = |\uparrow\uparrow\uparrow\uparrow\rangle,
\end{equation}
\begin{equation}
E_{2} = \frac{5}{4}J - \mu_{B}gh,
\end{equation}
\begin{equation}
|\psi_{2}\rangle = \frac{1}{2\sqrt{3}}\left(3|\downarrow\uparrow\uparrow\uparrow\rangle
-|\uparrow\downarrow\uparrow\uparrow\rangle
-|\uparrow\uparrow\downarrow\uparrow\rangle
-|\uparrow\uparrow\uparrow\downarrow\rangle\right).
\end{equation}

The corresponding critical field is
\begin{equation}
h_c = -\frac{2J}{\mu_B g} = 35.7212~\text{T}.
\end{equation}
The significantly larger critical field in the $Cu_5$ cluster is a result of the enhanced collective effects that arise from the greater interaction strength.

The complete isotropic spectrum follows by introducing the peripheral spin $\vec{S}_p=\sum_{i=2}^{N}\vec{S}_i$ and the total spin
$\vec{S}=\vec{S}_1+\vec{S}_p$. Using
$2\vec{S}_1\cdot\vec{S}_p=\vec{S}^2-\vec{S}_p^2-3/4$,
we obtain
\begin{equation}
E(s,S,M)=-\frac{\alpha J}{2}
\left[S(S+1)-s(s+1)-\frac{3}{4}\right]-g\mu_B hM,
\end{equation}
where $\alpha=1$ for $Cu_4$ and $\alpha=2$ for $Cu_5$.
For $n=N-1$ peripheral spins, $s=n/2,n/2-1,\ldots,s_{\min}$,
with $s_{\min}=0$ for even $n$ and $s_{\min}=1/2$ for odd $n$.
The allowed total spins are $S=s\pm1/2$ for $s>0$, and
$S=1/2$ for $s=0$, with $M=-S,-S+1,\ldots,S$. Each $(s,S,M)$ occurs with multiplicity
\begin{equation}
d_n(s)=\binom{n}{n/2-s}-\binom{n}{n/2-s-1},
\end{equation}
where binomial coefficients outside their allowed range vanish.

\begin{figure}[]
\centering
\includegraphics[width=0.48\textwidth]{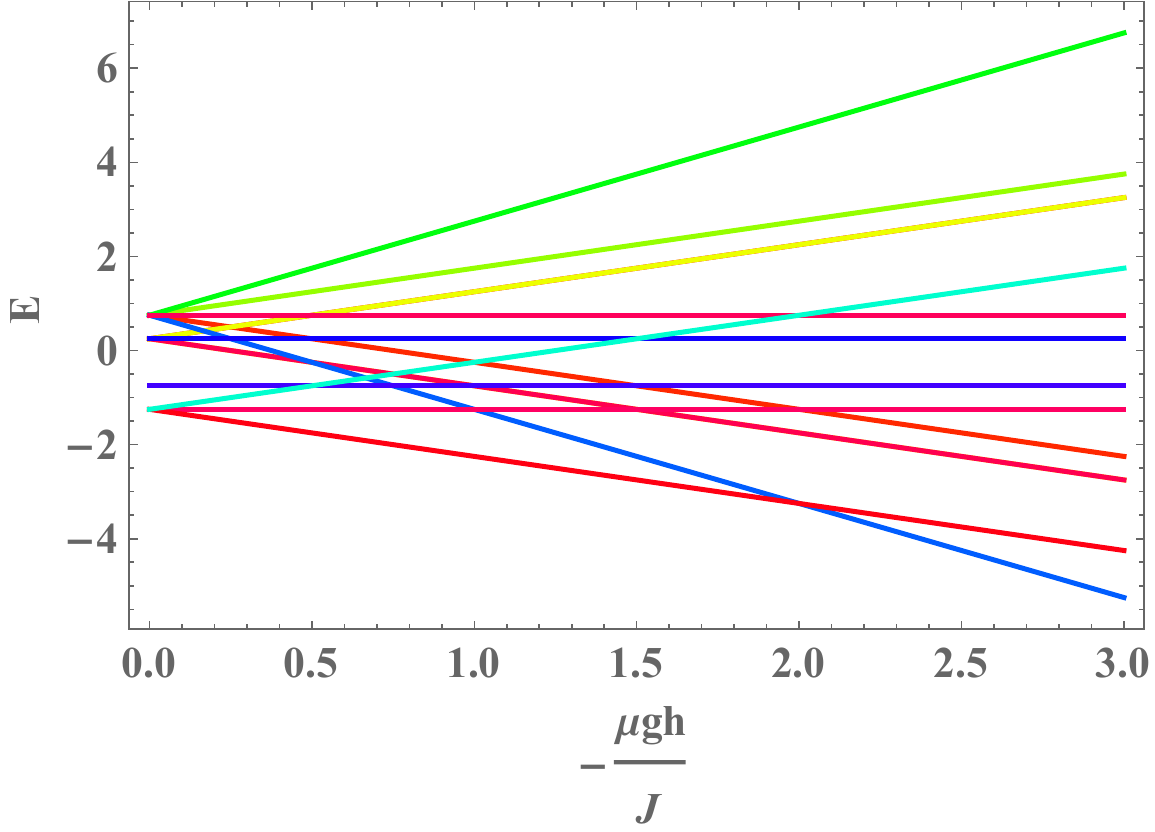}
\hfill
\includegraphics[width=0.48\textwidth]{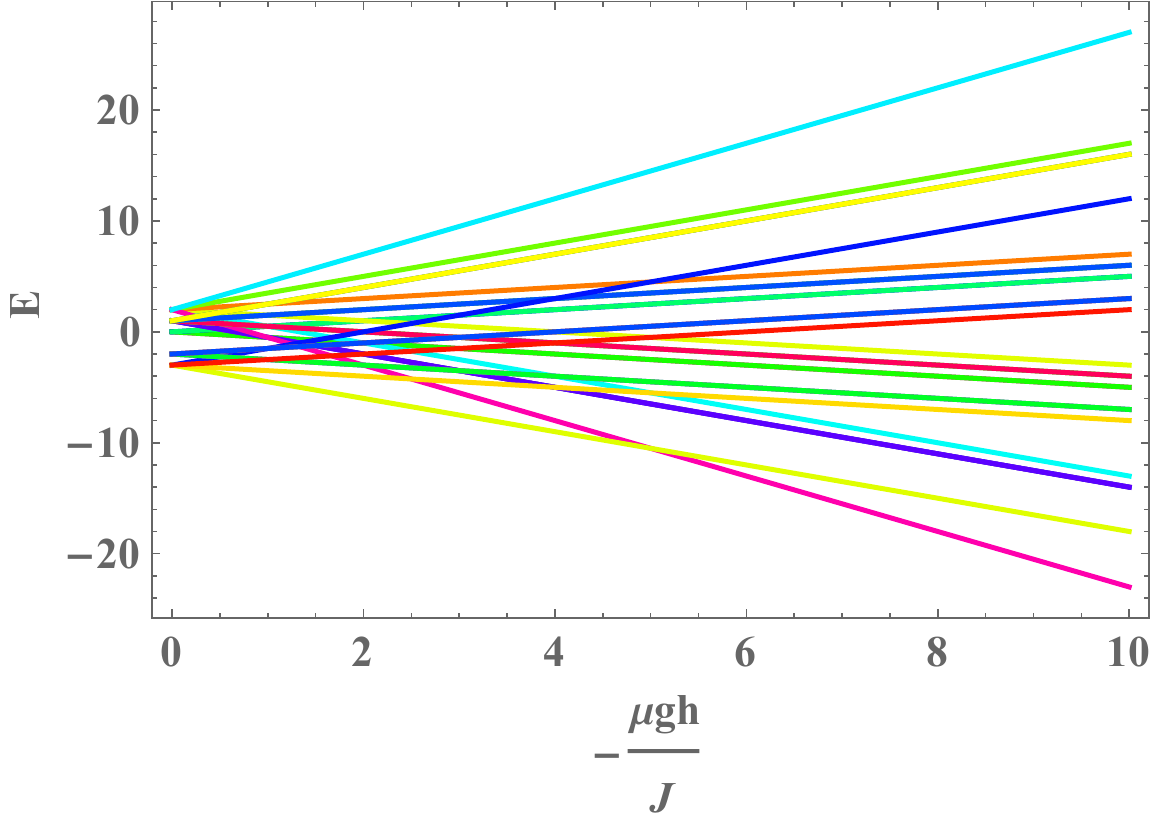}
\caption{Energy spectra of the copper clusters as a function of magnetic field: (left) the $Cu_4$ cluster and (right) the $Cu_5$ cluster.}
\label{fig:Cu_spectra}
\end{figure}

Figure \ref{fig:Cu_spectra} presents the corresponding energy spectra as functions of the magnetic field.

For antiferromagnetic exchange, the isotropic ground state has
the maximum peripheral spin $s=(N-1)/2$, coupled to the central
spin to give $S=s-1/2=(N-2)/2$. The ground state multiplets therefore
have $S=1$ for $Cu_4$ and $S=3/2$ for $Cu_5$. The unequal central
and peripheral sublattice spins leave an uncompensated moment,
giving these finite clusters a ferrimagnetic-like character without
implying macroscopic magnetic order. An applied field selects
the positive-$M$ components of these multiplets, before the
ground-state crossing into the fully polarized state at $h_c$.

From the partition function
\begin{equation}
\mathcal{Z} = \mathrm{Tr}\,[ e^{-\hat{H}/k_B T} ],
\end{equation}
the magnetization is defined as
\begin{equation}
M(T,h) = \frac{1}{\mathcal{Z}} \mathrm{Tr}\,\left[ \sum_{i=1}^{N} g \mu_B S_i^z \, e^{-\hat{H}/k_B T} \right],
\end{equation}
and the magnetic susceptibility is obtained via
\begin{equation}
\chi(T,h) = \frac{\partial M(T,h)}{\partial h}.
\end{equation}

These quantities provide a direct link to experimental observables, such as magnetization curves and temperature-dependent susceptibility \cite{Nayak2006, Chiang2016}. Quantum correlations are quantified through the entanglement of formation between spin pairs using the concurrence formalism \cite{Arnesen2001, Hill1997, Wootters1998}. For a reduced two-spin density matrix $\rho_r$, the concurrence is defined as
\begin{equation}
C = \max\{0, \lambda_1 - \lambda_2 - \lambda_3 - \lambda_4 \},
\end{equation}
where $\lambda_i$ are the square roots of the eigenvalues of $R = \rho_r (\sigma_y \otimes \sigma_y) \rho_r^* (\sigma_y \otimes \sigma_y)$ in descending order. The entanglement of formation is given by
\begin{equation}
E = -\frac{1+\sqrt{1-C^2}}{2} \log_2 \frac{1+\sqrt{1-C^2}}{2} - \frac{1-\sqrt{1-C^2}}{2} \log_2 \frac{1-\sqrt{1-C^2}}{2}.
\end{equation}

The thermal density matrix is expressed as
\begin{equation}
\rho(T) = \frac{e^{-\hat{H}/k_B T}}{\mathcal{Z}},
\label{ro}
\end{equation}
and the reduced density matrix $\rho_r$ is obtained by tracing out all degrees of freedom except the selected spin pair \cite{Arnesen2001,Baxter1982}. In star-shaped clusters, the central spin plays a crucial role in mediating correlations among peripheral spins. The resulting entanglement structure sensitively depends on the interplay between exchange interactions, magnetic field, and temperature. Thermal entanglement may be enhanced at intermediate temperatures, but vanishes beyond critical thresholds \cite{Arnesen2001}.

In the following sections, we extend this model by including the Dzyaloshinskii--Moriya interaction and analyze its influence on both magnetic and entanglement properties.

\section{Temperature Dependence of Magnetic Susceptibility}

We investigate the magnetic response of the tetranuclear $Cu_4$ and pentanuclear $Cu_5$ clusters through the temperature dependence of the magnetic susceptibility $\chi(T)$. The analysis is based on the antiferromagnetic Hamiltonians introduced in Section~2 (Eqs.~\ref{eq:hamCu5} and ~\ref{eq:hamCu4}). The product $T\chi(T)$ is particularly informative in identifying magnetic correlations and deviations from Curie behavior, providing insight into the underlying spin structure and exchange interactions \cite{Kahn1993, Nayak2006, Chiang2016, Papoyan2024,Carlin1986}.

The calculations are performed for a finite Dzyaloshinskii--Moriya interaction $G = 1\,\mathrm{cm}^{-1}$ and an external magnetic field $h = 0.5\,\mathrm{T}$ \cite{Ananikian2012, Rojas2012}. These theoretical results capture essential features observed in experiments on Cu(II) clusters, including the temperature-dependent suppression of magnetic susceptibility and the emergence of low-spin ground states. The qualitative analysis of the comparison with experimental data and their fit, presented in Fig. 6 of \cite{Nayak2006} and Fig. 4 of \cite{Chiang2016}, confirms the validity of the chosen Hamiltonian parameters and the role of Dzyaloshinskii–Moriya interactions in real molecular systems, as discussed in the next two Subsections.

\subsection{Tetranuclear \texorpdfstring{$Cu_4$}{Cu4} Cluster}

\begin{figure}[]
\centering
\includegraphics[width=0.6\textwidth]{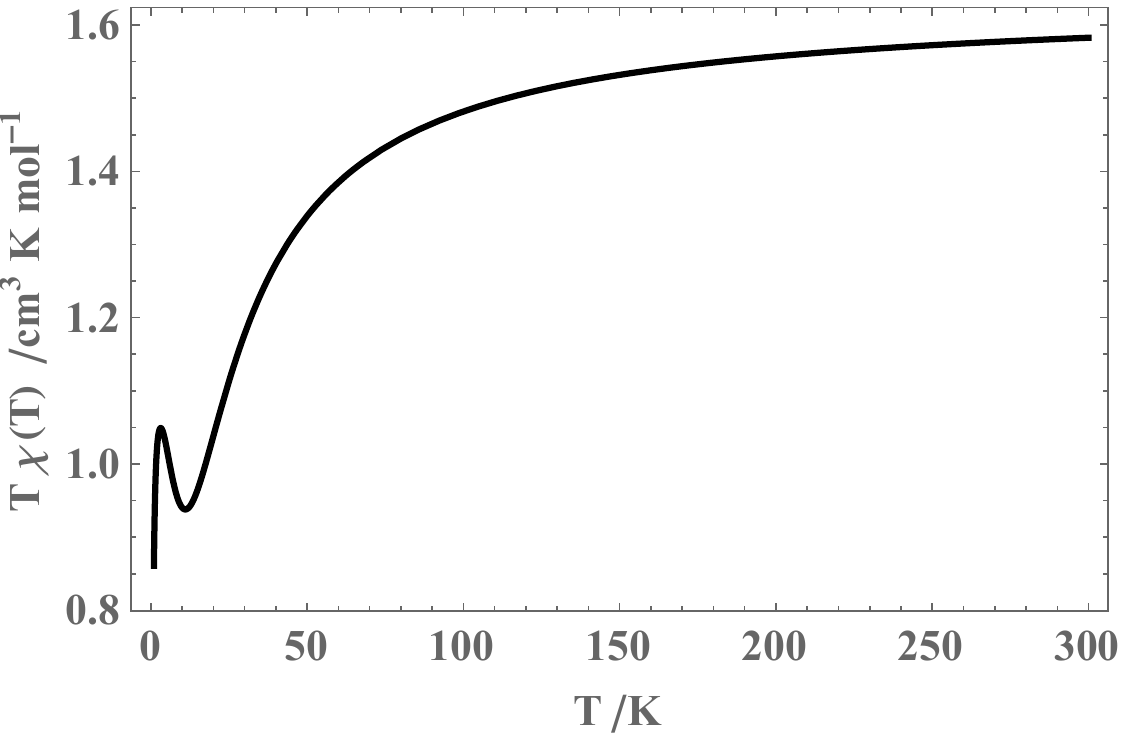}
\caption{Magnetic susceptibility $T\chi(T)$ as a function of temperature $T$ for the $Cu_4$ cluster at $h = 0.5\,\mathrm{T}$. The deviation from Curie behavior at low temperatures reflects antiferromagnetic correlations between the central and peripheral spins.}
\label{fig:Cu4_susceptibility_T}
\end{figure}

Figure~\ref{fig:Cu4_susceptibility_T} presents the temperature dependence of $T\chi(T)$ for the tetranuclear $Cu_4$ cluster.

At high temperatures, thermal fluctuations dominate and spins behave nearly independently, resulting in $T\chi(T)$ approaching the Curie law for paramagnets \cite{Kahn1993,Carlin1986}. As the temperature decreases, deviations from Curie behavior appear, indicating the formation of antiferromagnetically correlated states in which the central spin aligns antiparallel to the peripheral spins. 
However, the antiferromagnetic coupling does not fully compensate the cluster moment since the isotropic ground multiplet has $S=1$. On cooling, $T\chi(T)$ passes through a minimum near $11\,\mathrm{K}$ and subsequently increases as the excited $S=0$ singlets become depopulated. This upturn reflects the uncompensated ground state spin and the associated finite cluster ferrimagnetic-like character. Below the maximum near $3\,\mathrm{K}$, polarization of this moment by the finite applied field suppresses the susceptibility, producing the final downturn. The qualitative behavior of $T\chi(T)$ presented in Figure~\ref{fig:Cu4_susceptibility_T} is consistent with experimental observations for tetranuclear Cu(II) clusters \cite{Nayak2006}. 
A comparison between the results shown in Fig. ~\ref{fig:Cu4_susceptibility_T} and Fig. 6 of \cite{Nayak2006} shows that $T\chi(T)$ in the units of the $y$-axis of our Fig. ~\ref{fig:Cu4_susceptibility_T} is $\approx0.95$ at $T=20K$, while the corresponding experimental point is $\approx 0.60$. At $T=50,100,150,200,250$ our results for $T\chi(T)$ are respectively $1.34, 1.48, 1.53, 1.56, 1.57$ to be compared against the experimental values $1.21,1.45,1.49,1.51,1.52$, which were obtained at $h=1\,\mathrm{T}$.

\subsection{Pentanuclear \texorpdfstring{$Cu_5$}{Cu5} Cluster}

\begin{figure}[]
\centering
\includegraphics[width=0.6\textwidth]{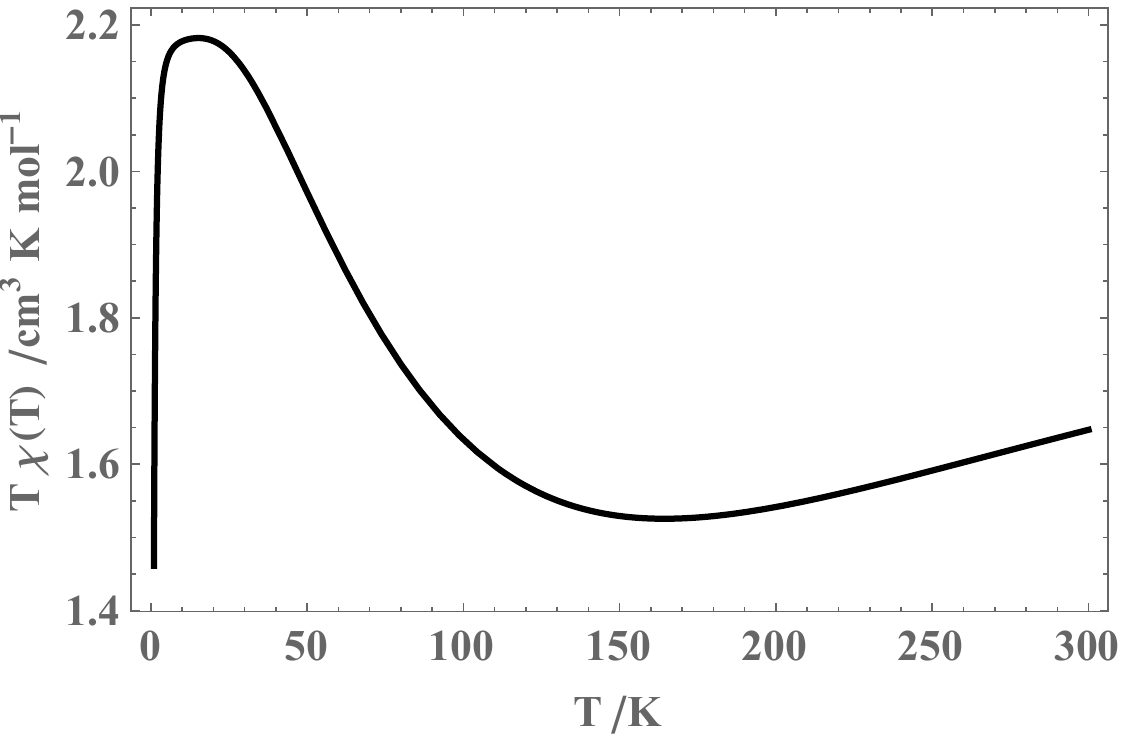}
\caption{Magnetic susceptibility $T\chi(T)$ as a function of temperature $T$ for the $Cu_5$ cluster at $h=0.5\,\mathrm{T}$. The stronger deviation from Curie behavior reflects enhanced collective antiferromagnetic correlations due to the increased number of interacting spins.}
\label{fig:Cu5_susceptibility_T}
\end{figure}

Figure~\ref{fig:Cu5_susceptibility_T} shows the corresponding temperature dependence for the pentanuclear $Cu_5$ cluster.

Compared to the $Cu_4$ cluster, the $Cu_5$ system exhibits a more pronounced deviation from Curie behavior at low temperatures. The central spin mediates interactions among the four peripheral spins, enhancing collective antiferromagnetic correlations and producing a richer spectrum of low-energy states \cite{Chiang2016, Papoyan2024, Chandrasekhar2009}. More specifically, on cooling below the broad minimum near
$164\,\mathrm{K}$, $T\chi(T)$ increases markedly as the
$S=3/2$ ground multiplet becomes dominant. In the isotropic
spectrum, the lowest excited multiplets have $S=1/2$. Their depopulation restores the larger response of the
uncompensated ground state spin. The maximum near $15\,\mathrm{K}$ is followed by a downturn at the lowest temperatures due to field-induced polarization within the ground state multiplet.
As the temperature increases, thermal excitations populate higher-energy states, gradually restoring paramagnetic behavior \cite{Kahn1993, Carlin1986}. 

These results are consistent with experimental measurements on pentanuclear Cu(II) complexes \cite{Chiang2016, Papoyan2024, Mondal2019}, demonstrating that cluster geometry and exchange interactions govern the magnetic response \cite{Chandrasekhar2009, Bhowmika2018}. More specifically, a comparison between the results shown in Fig. ~\ref{fig:Cu5_susceptibility_T} and Fig. 4 of \cite{Chiang2016} shows that $T\chi(T)$ in the units of the $y$-axis of our Fig. ~\ref{fig:Cu5_susceptibility_T} is $\approx 2.18$ at $T=20K$, while the corresponding experimental point is $\approx2.08$. At $T=50,100,150,200,250$ our results for $T\chi(T)$ are respectively $1.97, 1.63, 1.53, 1.54, 1.59$ to be compared against the experimental values $1.95,1.62,1.53,1.55,1.60$, which were obtained at $h=0.1\,\mathrm{T}$.

\subsection{Discussion}

The temperature dependence of $T\chi(T)$ reveals clear signatures of antiferromagnetic coupling in both clusters. The star-shaped topology, from the results we presented, significantly influences the ground states and produces characteristic deviations from the Curie law at low temperatures \cite{Carlin1986, Kahn1993, Chandrasekhar2009}. The comparison between $Cu_4$ and $Cu_5$ highlights the effect of cluster size: Increasing the number of peripheral spins enhances collective effects and strengthens magnetic correlations \cite{Bhowmika2018, Mondal2019}. However, we are unable to clearly demonstrate this effect due to the large differences in the exchange interaction values for $Cu_4$ and $Cu_5$. These findings provide a coherent picture of the magnetic behavior of homometallic Cu(II) clusters and support the applicability of the proposed theoretical model.

\section{Entanglement Entropy}

Quantifying entanglement in finite-size clusters provides deeper insight into quantum correlations beyond conventional magnetic observables. The spatial distribution of entanglement and its sensitivity to magnetic field and anisotropic interactions reveal how quantum information is stored and redistributed within molecular spin networks. These insights are relevant for potential applications in quantum computing, quantum sensing, and molecular spintronics. Here and below, EE stands for entanglement entropy between nearest-neighbor pairs.

We analyze quantum correlations in the \texorpdfstring{$Cu_4$}{Cu4} and \texorpdfstring{$Cu_5$}{Cu5} star-shaped clusters using the entanglement of formation, computed from the thermal density matrix defined in Eq.~(\ref{ro}) \cite{Rojas2025, Hill1997, Wootters1998}. The dependence of entanglement on magnetic field $h$, temperature $T$, and Dzyaloshinskii--Moriya interaction $G$ is discussed below in connection with the calculated figures \cite{Rojas2025, Papoyan2024}.

%-------------------------------------------------
\subsection{Field dependence of entanglement, magnetization, and susceptibility}

\begin{figure}[]
\centering
\includegraphics[width=0.45\textwidth]{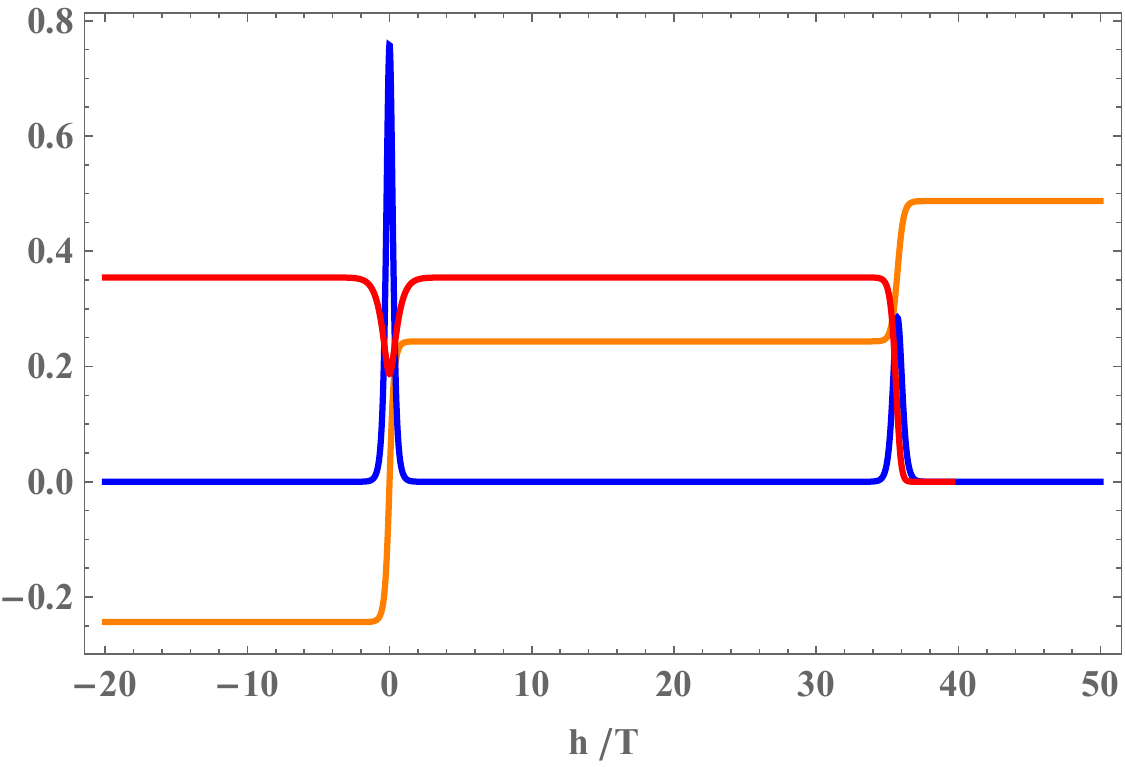}
\includegraphics[width=0.45\textwidth]{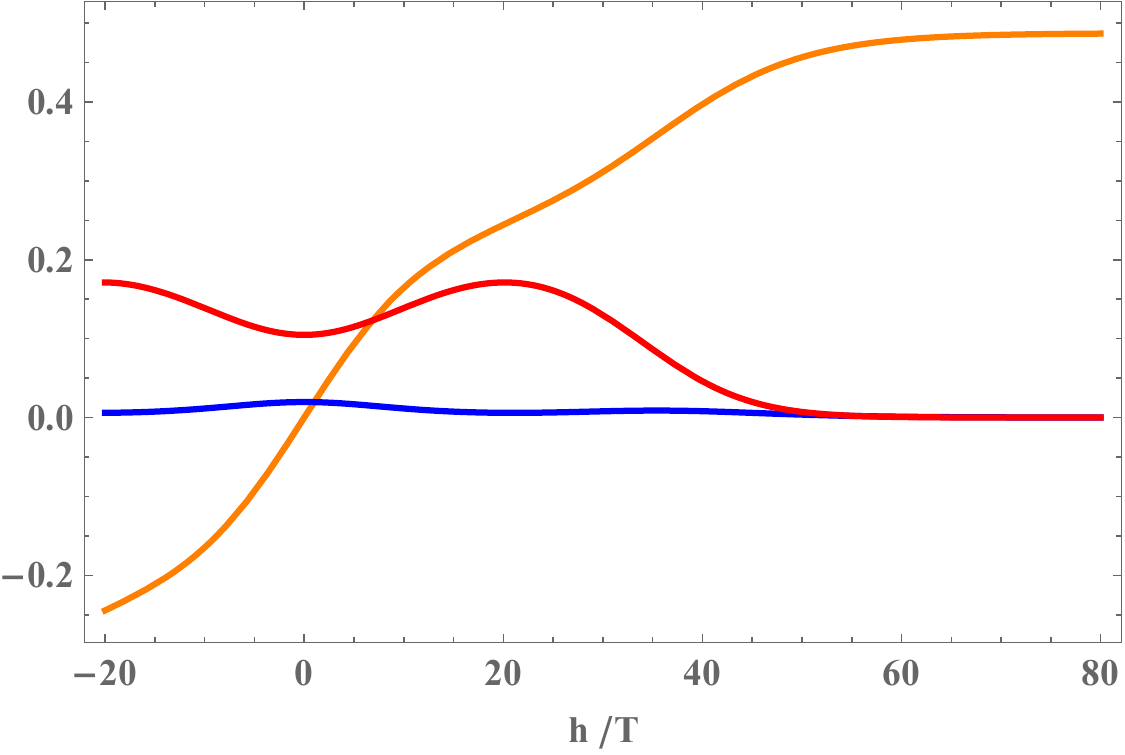}
\caption{$Cu_4$ cluster: Entanglement entropy $EE$ (red), magnetization $M$ (orange), and magnetic susceptibility $\chi$ (blue) as functions of magnetic field $h$  for (left) $T=0.3$ K and (right) $T=10$ K.}
\label{fig:Cu4_eemh}
\end{figure}

\begin{figure}[]
\centering
\includegraphics[width=0.45\textwidth]{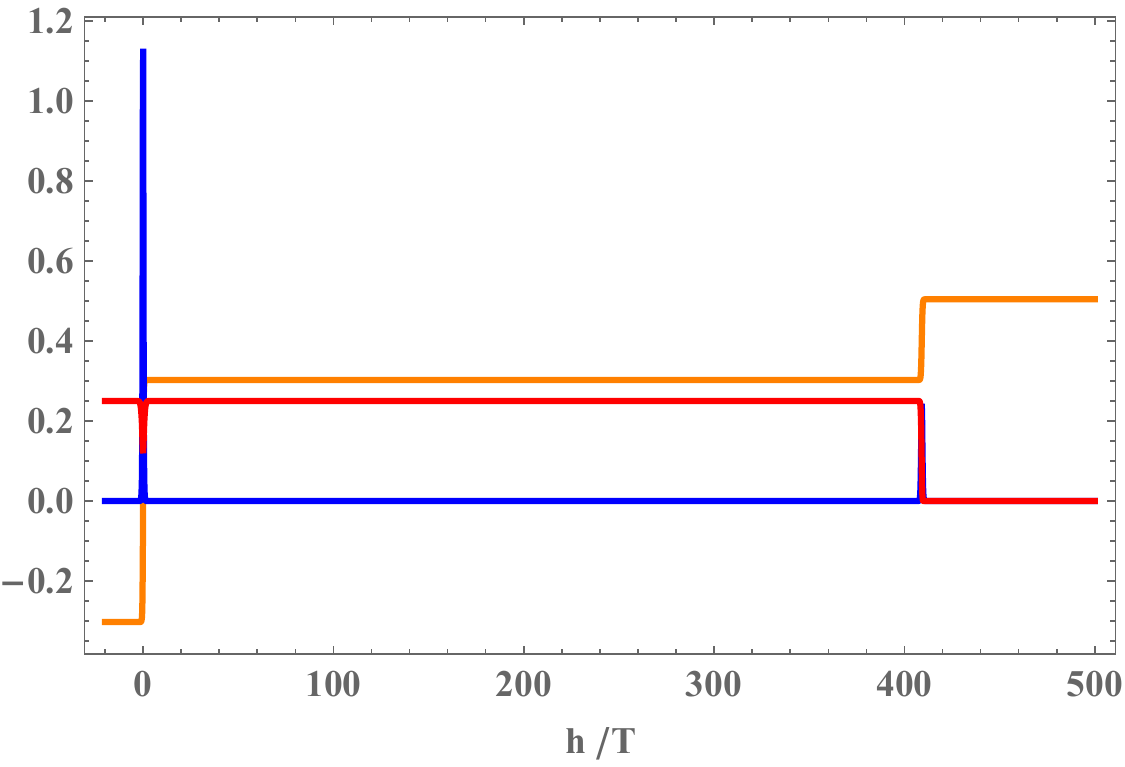}
\includegraphics[width=0.45\textwidth]{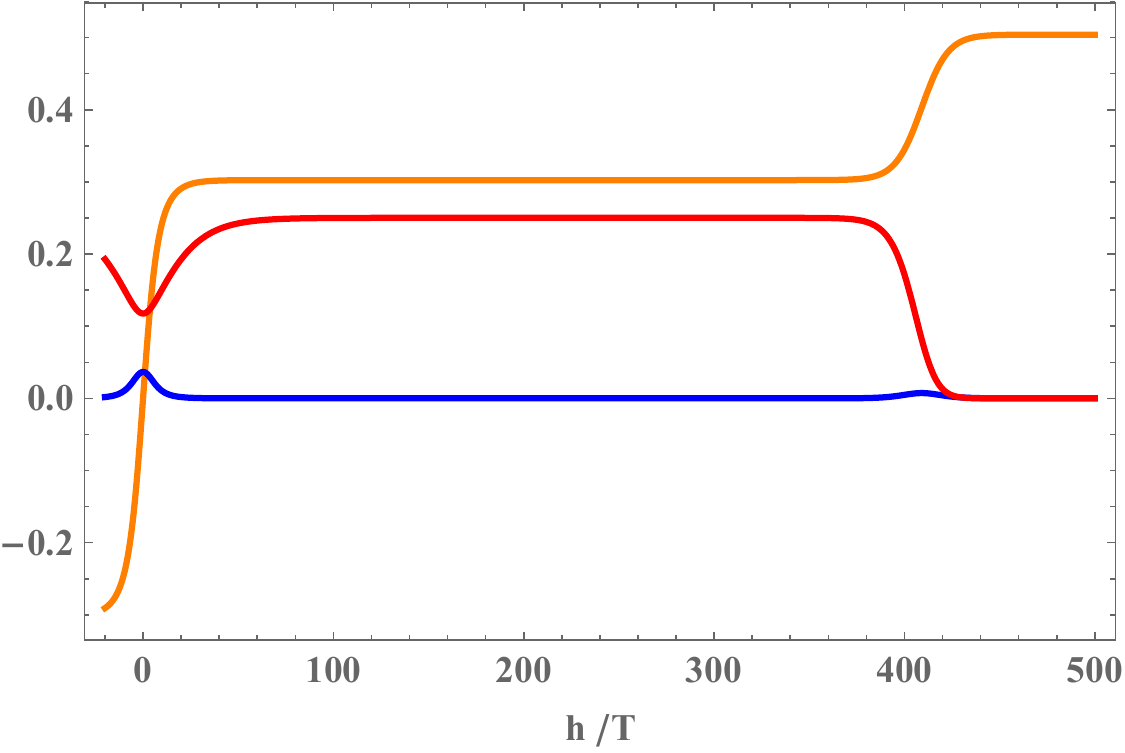}
\caption{$Cu_5$ cluster: Entanglement entropy $EE$ (red), magnetization $M$ (orange), and magnetic susceptibility $\chi$ (blue) as functions of magnetic field $h$  for (left) $T=0.3$ K and (right) $T=10$ K.}
\label{fig:Cu5_eemh}
\end{figure}

As shown in Figs.~\ref{fig:Cu4_eemh} and \ref{fig:Cu5_eemh}, the entanglement exhibits pronounced variations as a function of the magnetic field. At low temperature ($T=0.3$ K), sharp features emerge: peaks in $\chi$ coincide with rapid changes in $M$, indicating field-induced level crossings. Importantly, the entanglement entropy attains local maxima near these transitions, demonstrating that quantum correlations are strongest in regimes where competing spin configurations coexist \cite{Arnesen2001, Amico2008, Eisert2010}.

At higher temperature ($T=10$ K), thermal fluctuations smooth these features, leading to a substantial reduction of entanglement and a more gradual field dependence of both magnetization and susceptibility \cite{Eisert2010, Girolami2013}.

%-------------------------------------------------
\subsection{Nearest- and next-nearest-neighbor entanglement}

\begin{figure}[]
\centering
\includegraphics[width=0.45\textwidth]{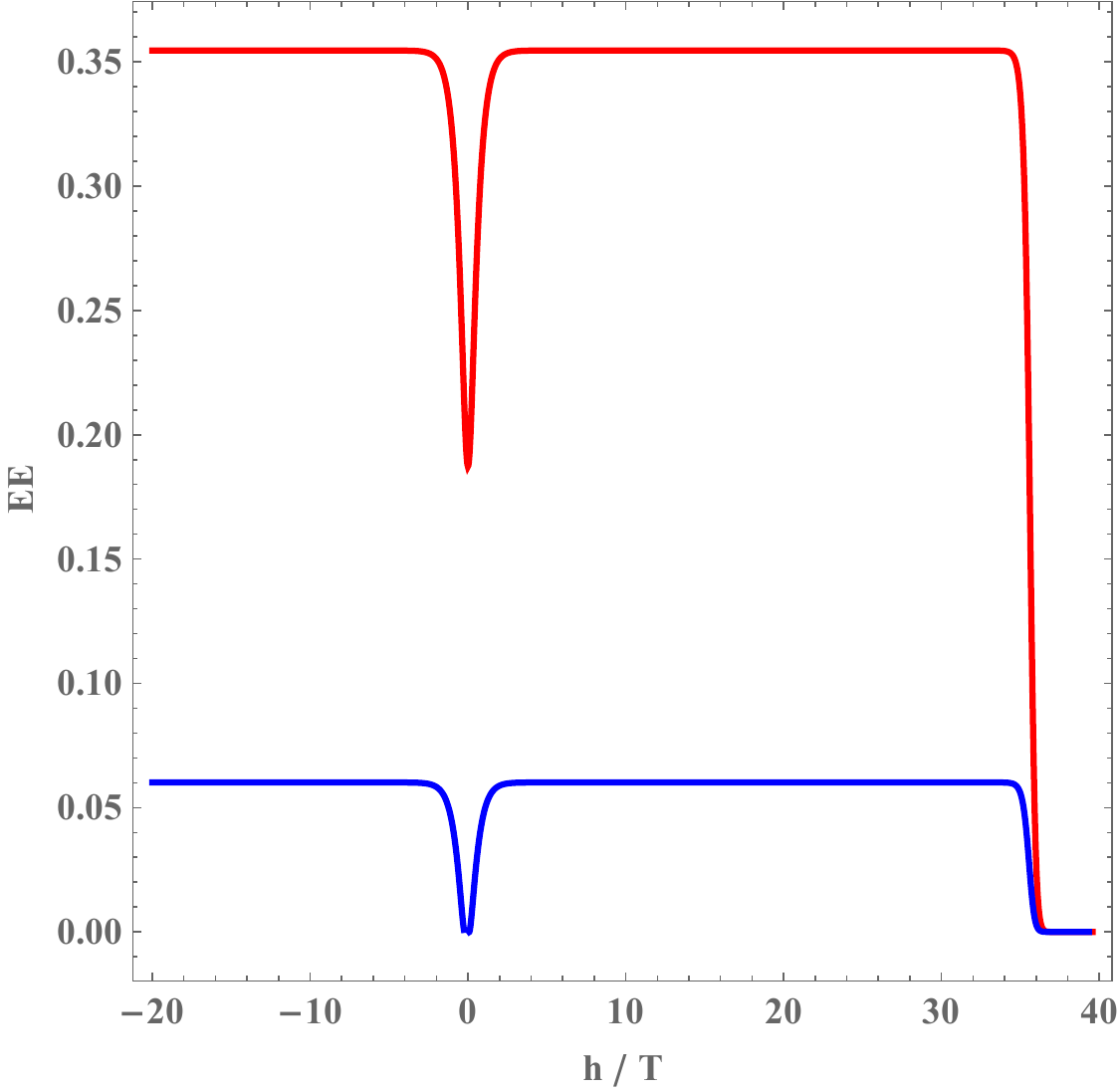}
\includegraphics[width=0.45\textwidth]{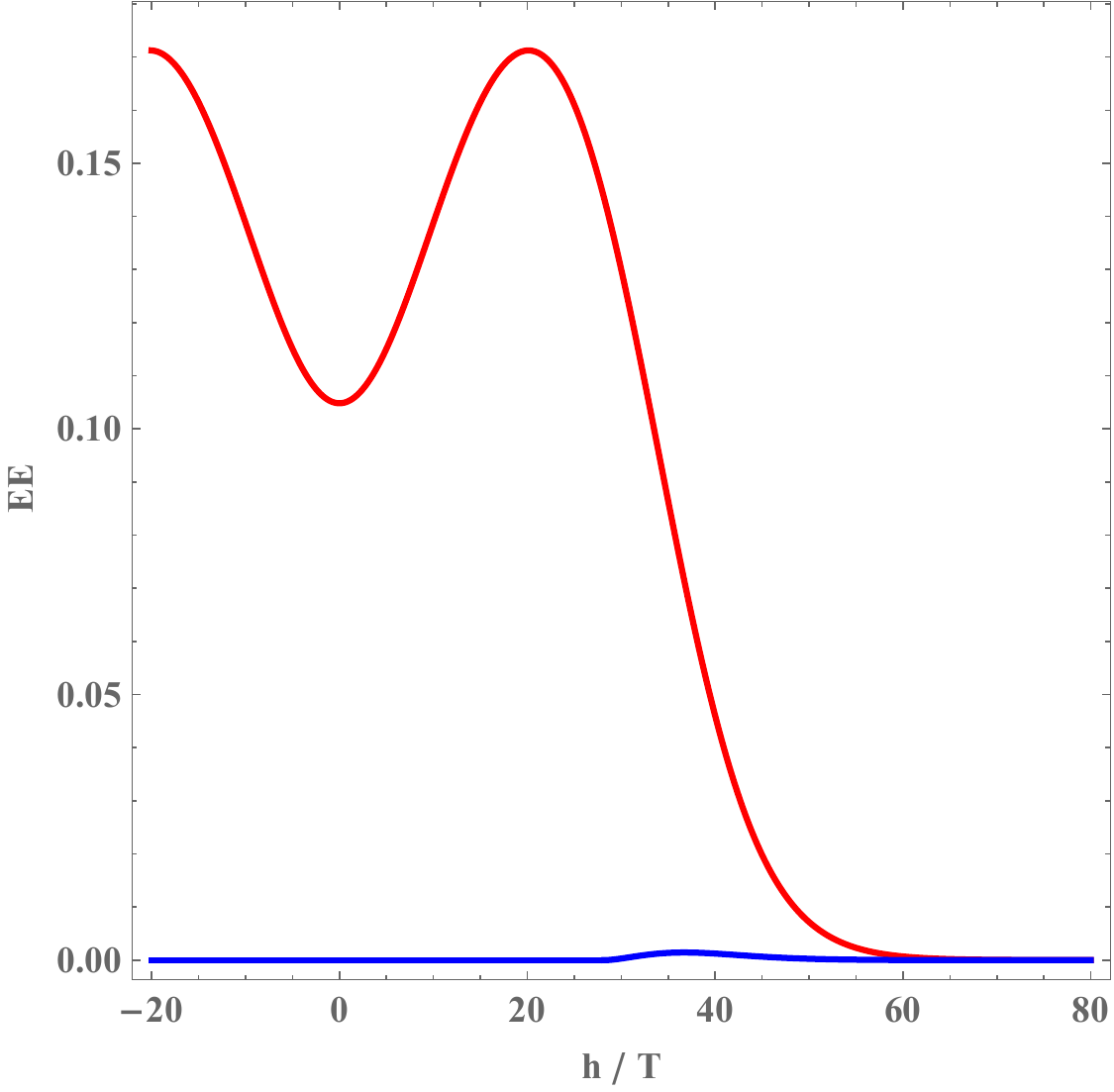}
\caption{$Cu_4$ cluster: Entanglement entropy of nearest neighbors $EE_{12}=EE_{13}=EE_{14}$ (red curve) and next-nearest neighbors $EE_{23}=EE_{24}=EE_{34}$ (blue curve) as a function of magnetic field $h$ for (left) $T=0.3$ K and (right) $T=10$ K.}
\label{fig:Cu4_pair}
\end{figure}

\begin{figure}[]
\centering
\includegraphics[width=0.45\textwidth]{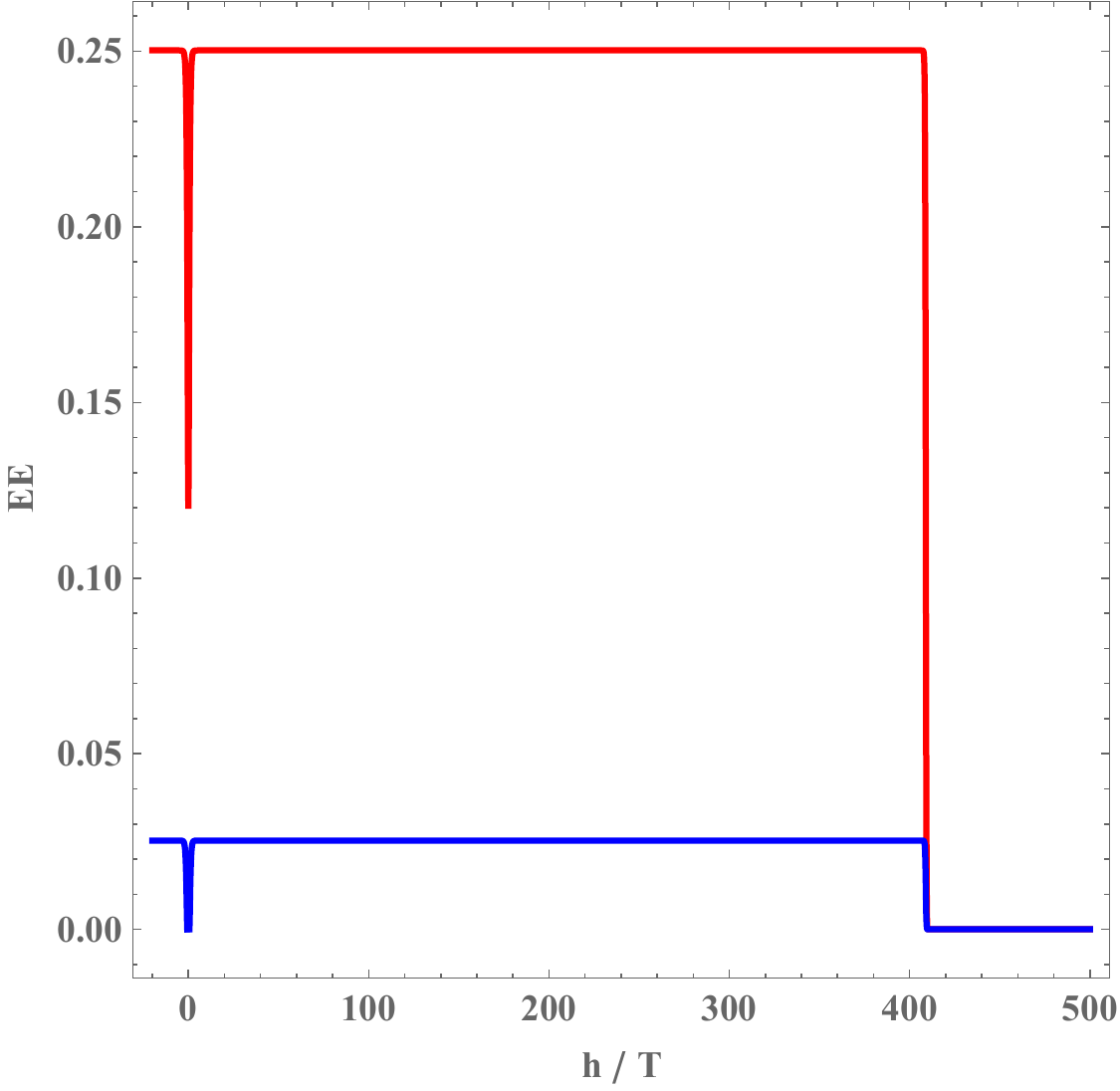}
\includegraphics[width=0.45\textwidth]{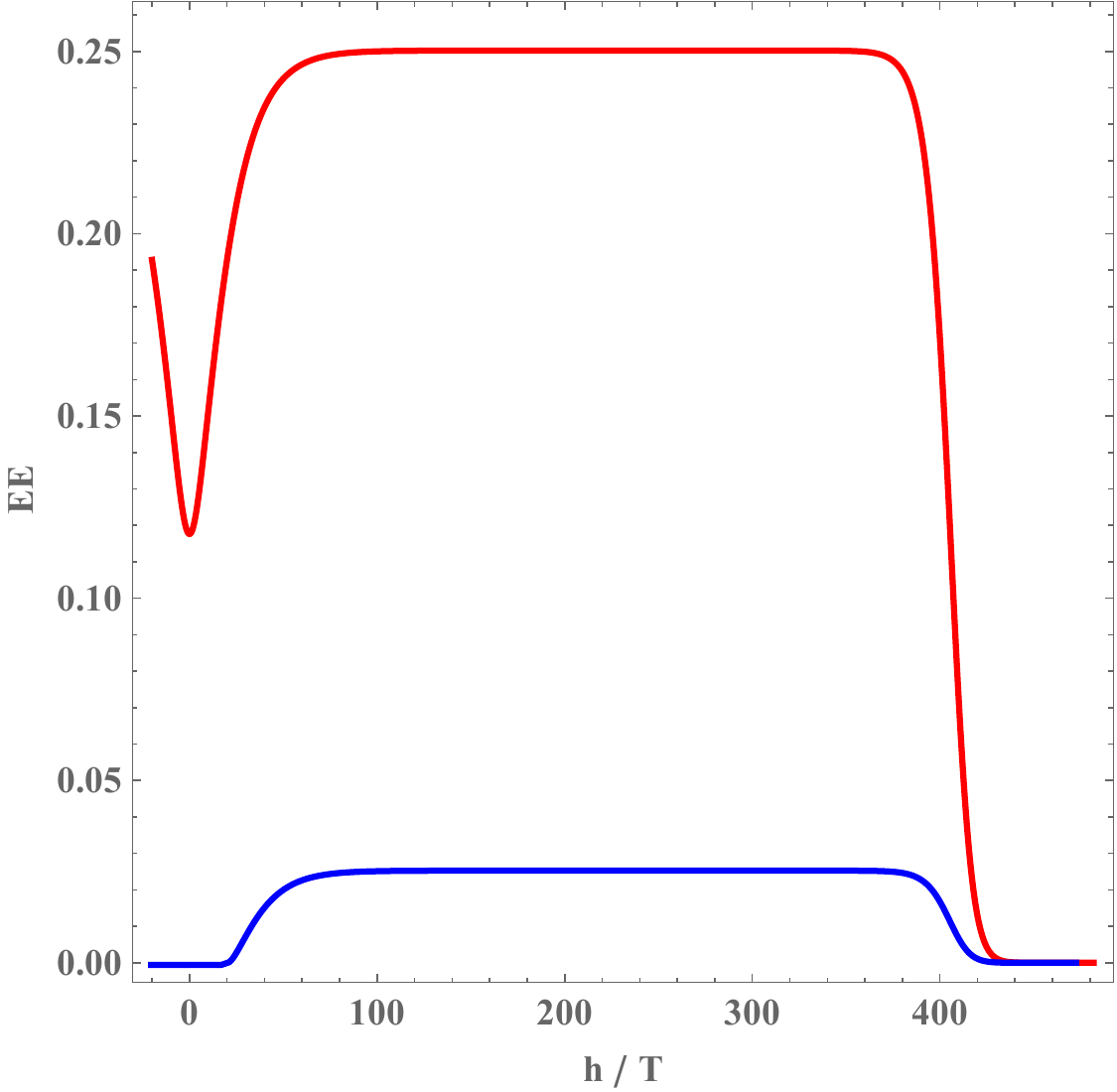}
\caption{$Cu_5$ cluster: Entanglement entropy of nearest neighbors $EE_{12}=EE_{13}=EE_{14}=EE_{15}$ (red curve) and next-nearest neighbors $EE_{23}=EE_{24}=EE_{25}=EE_{34}=EE_{35}=EE_{45}$ (blue curve) as a function of magnetic field $h$ for (left) $T=0.3$ K and (right) $T=10$ K.}
\label{fig:Cu5_pair}
\end{figure}

Figures~\ref{fig:Cu4_pair} and \ref{fig:Cu5_pair} reveal the spatial structure of the entanglement. At low temperature, nearest-neighbor correlations (central–peripheral spins) dominate, reflecting direct antiferromagnetic exchange interactions \cite{Kahn1993, Carlin1986, Chandrasekhar2009}. In contrast, next-nearest-neighbor entanglement (between peripheral spins) is weaker and arises indirectly through the central spin \cite{Nayak2006, Mondal2019}.

Near critical magnetic fields, both contributions become comparable, indicating a redistribution and delocalization of quantum correlations across the cluster \cite{Amico2008, Eisert2010}.

%-------------------------------------------------
\subsection{Effect of Dzyaloshinskii--Moriya interaction}
For consistency with experimental realizations, we begin with the isotropic Heisenberg model and then extend the analysis by including the antisymmetric Dzyaloshinskii--Moriya (DM) interaction. This interaction originates from spin--orbit coupling in systems lacking inversion symmetry at the exchange bonds and introduces anisotropic exchange effects that significantly influence the magnetic and entanglement properties of the clusters \cite{Dzyaloshinsky1958,Moriya1960}.

If we take into account the antisymmetric Dzyaloshinskii--Moriya (DM) interaction, an additional term is included in the Hamiltonian. For the considered star-shaped $Cu_N$ clusters, this contribution is written as
\begin{equation}
H_{\mathrm{DM}} = G \sum_{i=2}^{N} \left( S_{1}^{x} S_{i}^{y} - S_{1}^{y} S_{i}^{x} \right).
\end{equation}
Each DM vector is taken along the $\hat{z}$-direction, which is perpendicular to the $Cu_N$ plane, and has the same amplitude for all bonds. The parameter $G$ characterizes the strength of the Dzyaloshinskii--Moriya interaction, which introduces anisotropy, spin canting, and weak breaking of the spin-rotational symmetry.

\begin{figure}[]
\centering
\includegraphics[width=0.45\textwidth]{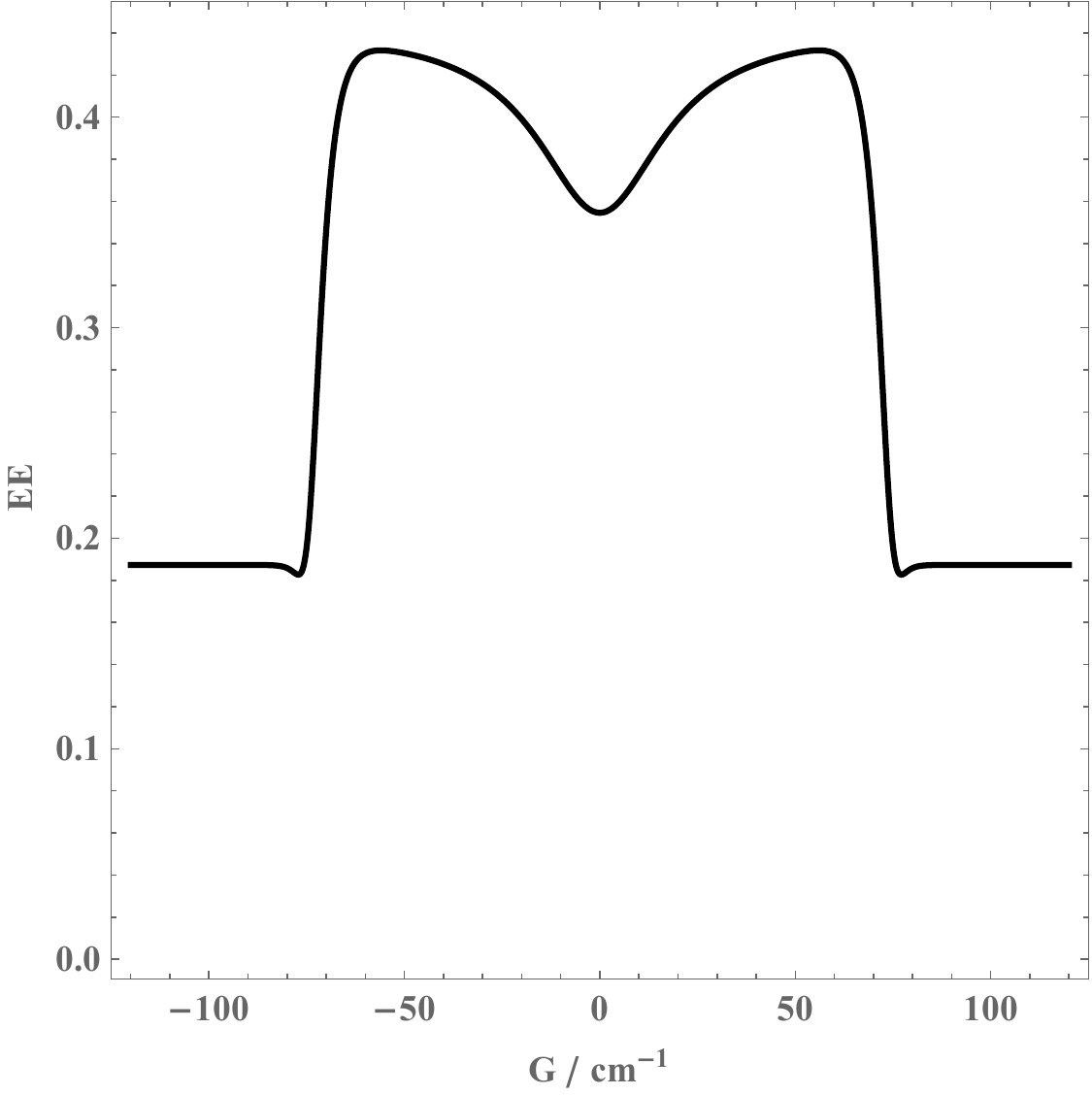}
\includegraphics[width=0.45\textwidth]{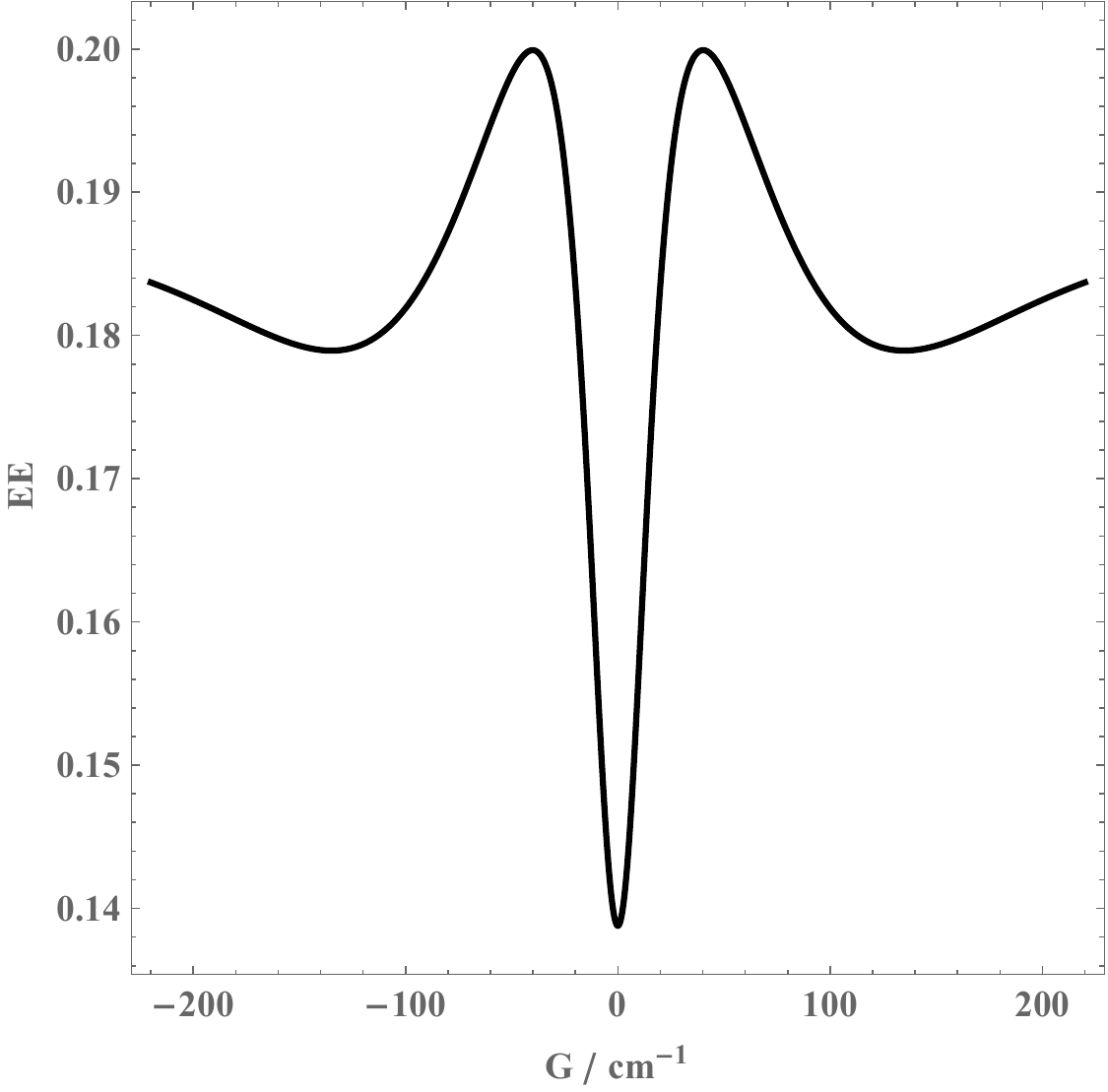}
\caption{$Cu_4$ cluster: Entanglement entropy as a function of the Dzyaloshinskii--Moriya interaction $G$ at $h=10$ T for (left) $T=0.3$ K and (right) $T=10$ K.}
\label{fig:Cu4_G}
\end{figure}

\begin{figure}[]
\centering
\includegraphics[width=0.45\textwidth]{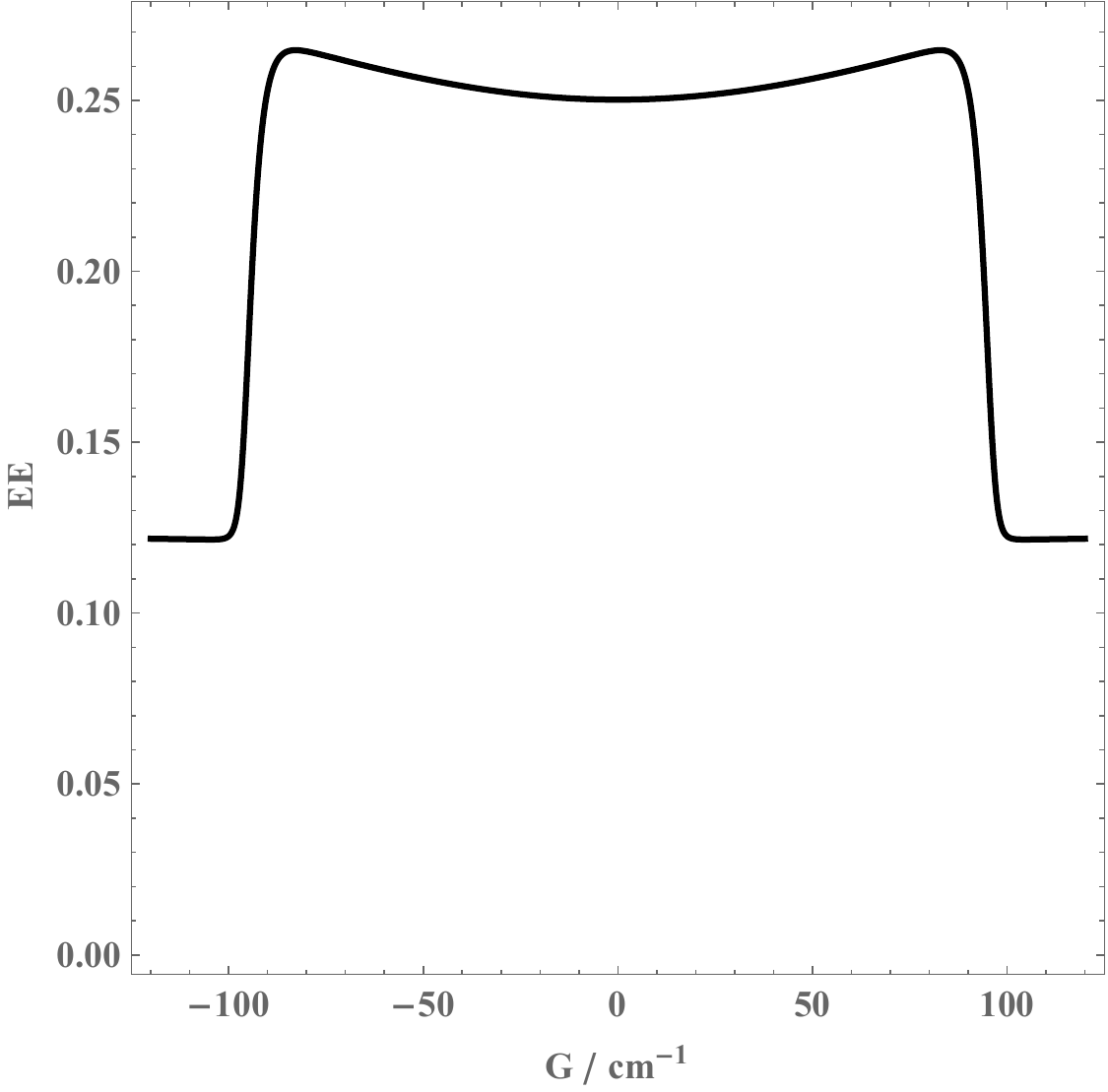}
\includegraphics[width=0.45\textwidth]{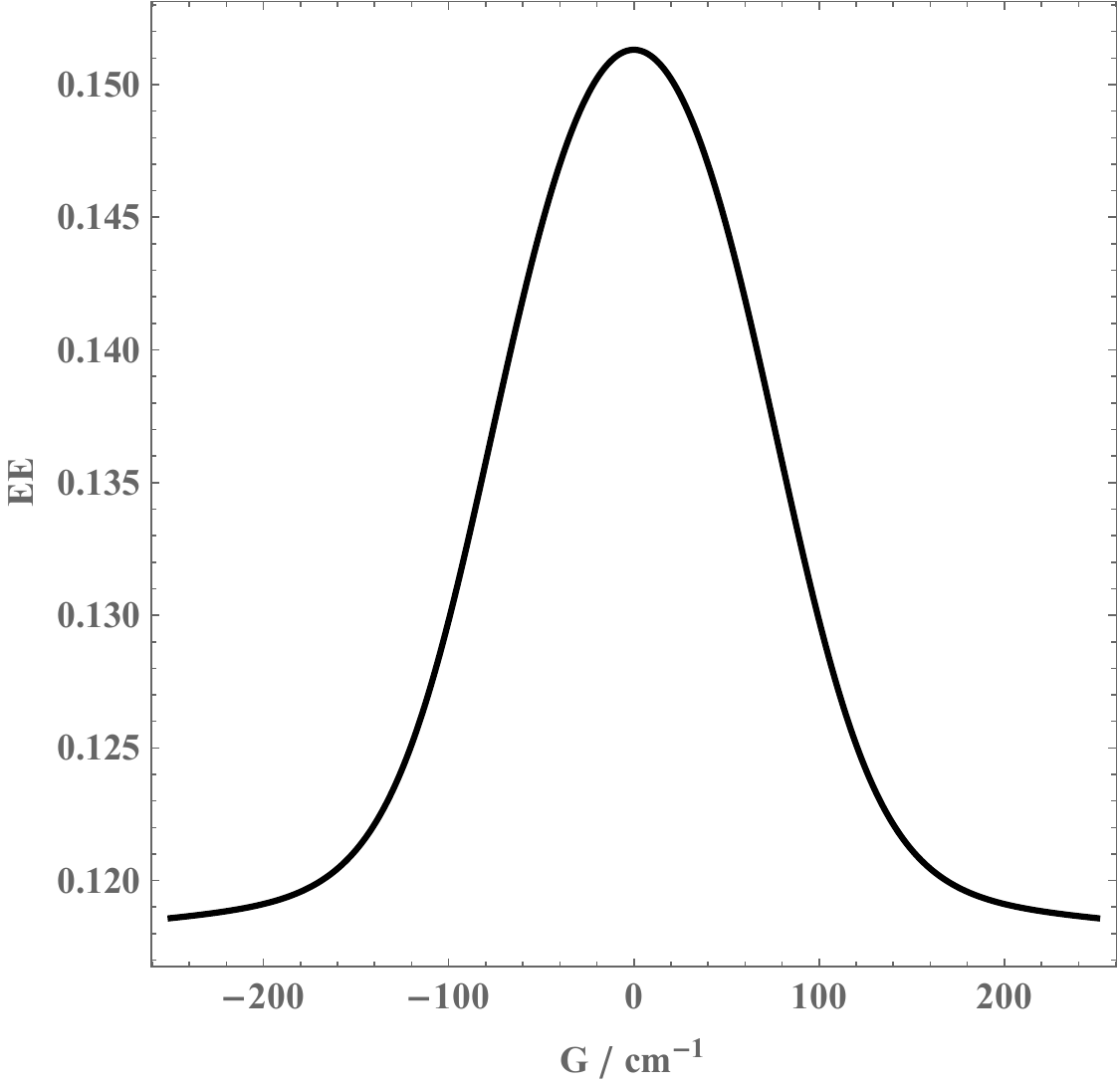}
\caption{$Cu_5$ cluster: Entanglement entropy as a function of the Dzyaloshinskii--Moriya interaction $G$ at $h=10$ T for (left) $T=0.3$ K and (right) $T=10$ K.}
\label{fig:Cu5_G}
\end{figure}

The Dzyaloshinskii--Moriya interaction enhances entanglement in the weak-$G$ regime by inducing spin canting and mixing of quantum states. For larger $G$, entanglement tends to saturate or decrease due to the dominance of anisotropic exchange effects \cite{Li2012}.

%-------------------------------------------------
\subsection{Global entanglement behavior in parameter space}

\begin{figure}[]
\centering
\includegraphics[width=0.45\textwidth]{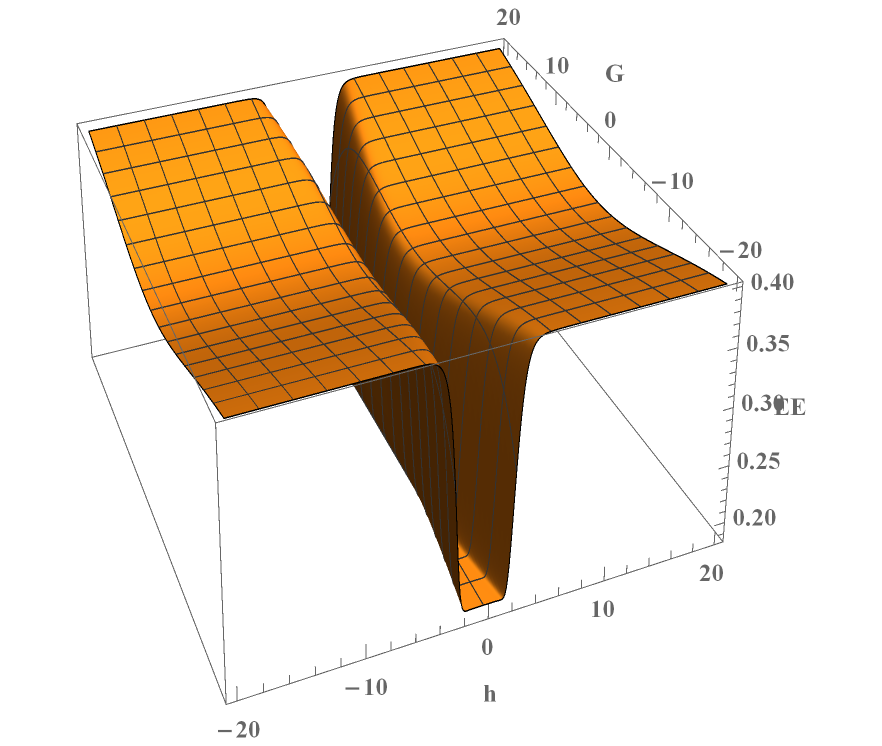}
\includegraphics[width=0.45\textwidth]{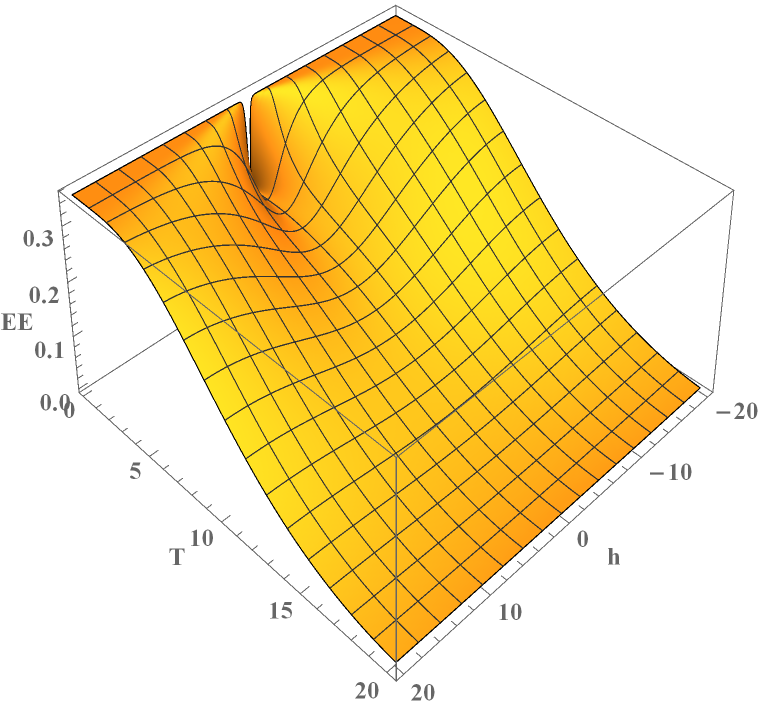}
\caption{$Cu_4$ cluster: Entanglement entropy as a function of (left) magnetic field $h$ and Dzyaloshinskii--Moriya interaction $G$ at $T=0.3$ K, and (right) magnetic field $h$ and temperature $T$ at $G=1\,\mathrm{cm}^{-1}$.}
\label{fig:Cu4_3D}
\end{figure}

\begin{figure}[]
\centering
\includegraphics[width=0.45\textwidth]{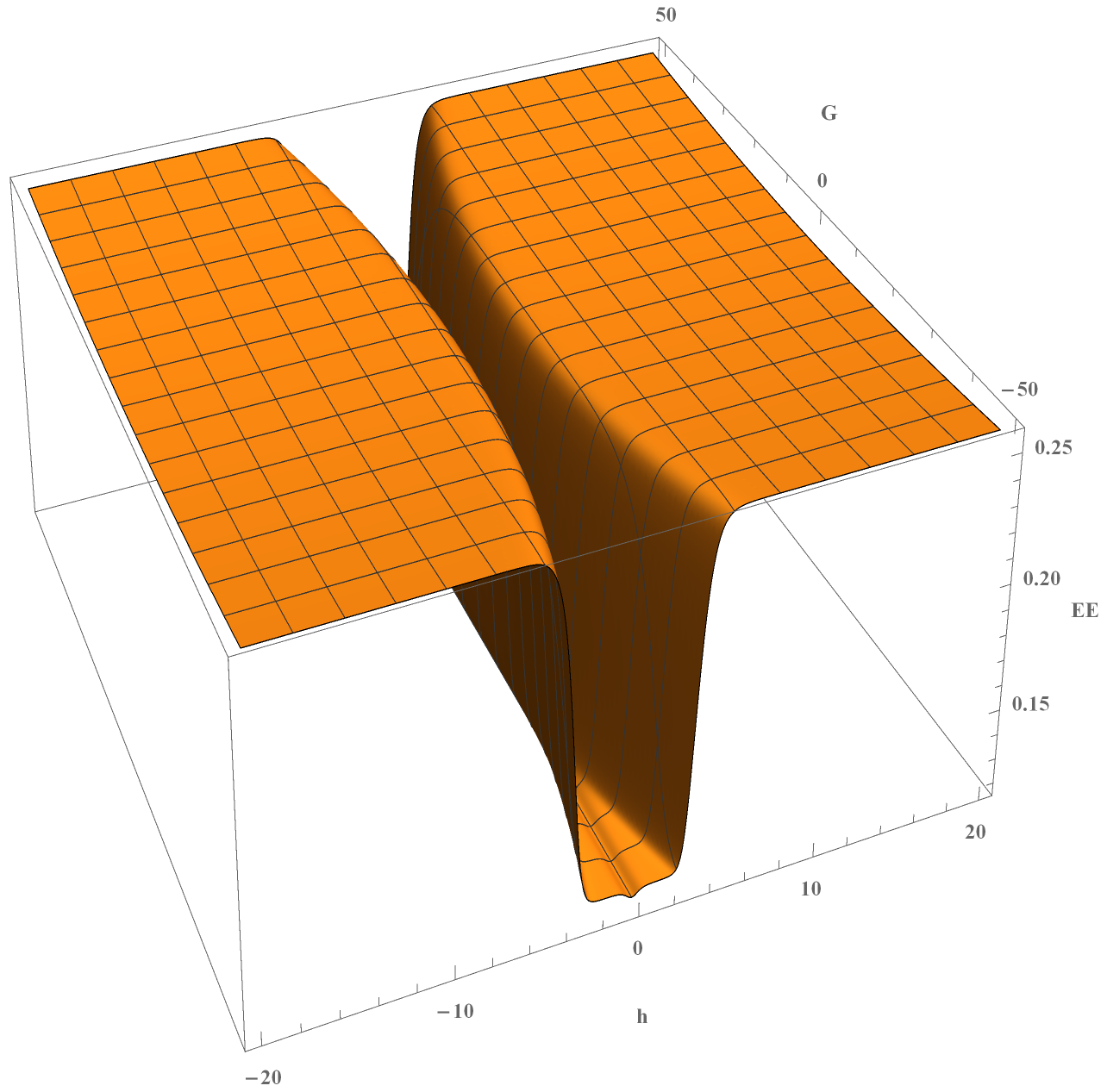}
\includegraphics[width=0.45\textwidth]{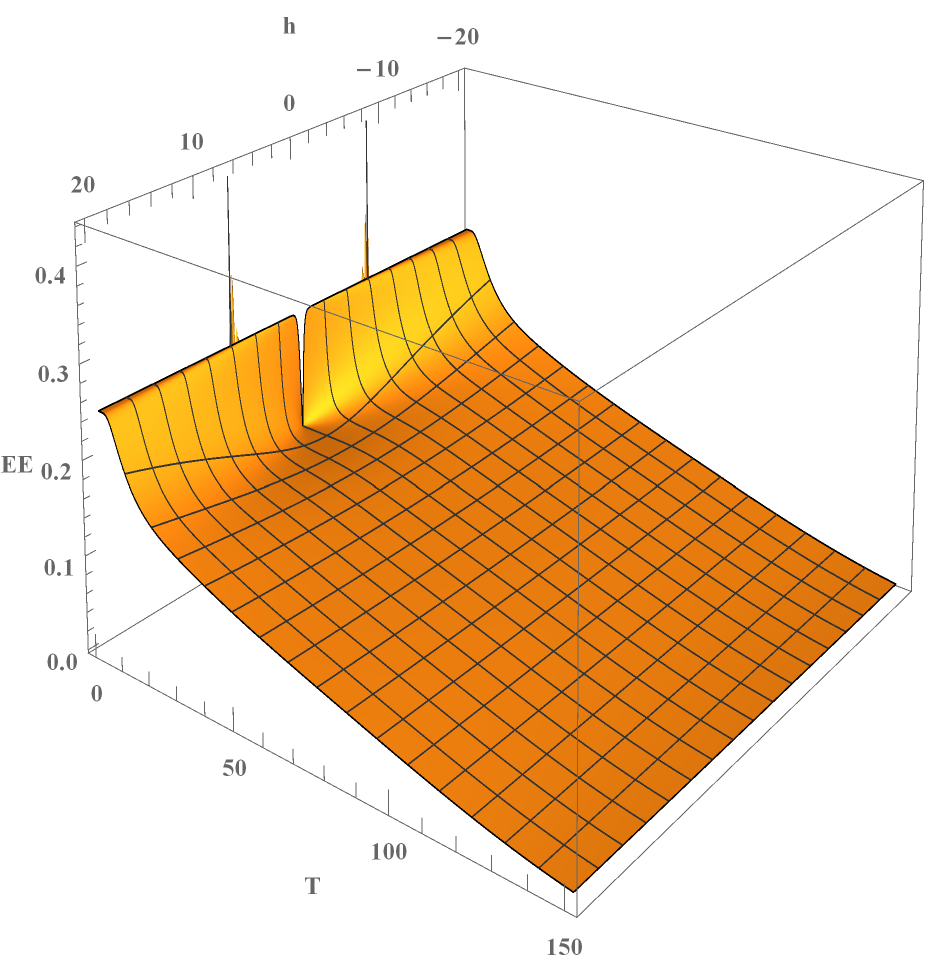}
\caption{$Cu_5$ cluster: Entanglement entropy as a function of (left) magnetic field $h$ and Dzyaloshinskii--Moriya interaction $G$ at $T=0.3$ K, and (right) magnetic field $h$ and temperature $T$ at $G=1\,\mathrm{cm}^{-1}$.}
\label{fig:Cu5_3D}
\end{figure}

Figures~\ref{fig:Cu4_3D} and \ref{fig:Cu5_3D} provide a comprehensive visualization of the entanglement across the parameter space. The left panels show that entanglement is strongly localized in specific regions of $(h,G)$ space, corresponding to magnetic-field-driven spin-state transitions. The dependence on \(G\) at fixed \(h=10\) T is shown in Figs.~\ref{fig:Cu4_G}  and~\ref{fig:Cu5_G}.

The right panels of Figs.~\ref{fig:Cu4_3D} and \ref{fig:Cu5_3D} illustrate the combined effects of temperature and magnetic field: entanglement is maximized at low temperatures and intermediate fields, while it rapidly decreases with increasing temperature due to thermal decoherence. In the presence of the
Dzyaloshinskii--Moriya interaction, the critical value of the magnetic field is slightly shifted; from exact analytical
calculations, we obtain $h_c = 409.112\,\mathrm{T}$  for the $Cu_5$  complex and $h_c = 35.703\,\mathrm{T}$ for the $Cu_4$ complex.

%-------------------------------------------------
\subsection{Discussion}

The results demonstrate that the entanglement in the $Cu_4$ and $Cu_5$ clusters is highly sensitive to external control parameters. The magnetic field induces sharp transitions associated with enhanced quantum correlations, whereas temperature suppresses entanglement. The Dzyaloshinskii--Moriya interaction provides an additional tuning mechanism, modifying both the magnitude and spatial distribution of the entanglement.

\section{Conclusion}

In this work, we have presented a theoretical 
study of the magnetic and entanglement entropy properties of star-shaped tetranuclear $Cu_4$ and pentanuclear $Cu_5$ clusters within the framework of the antiferromagnetic Heisenberg model. The analysis was motivated by experimentally realized molecular systems and focused on the interplay between magnetic observables and quantum correlations.

A detailed investigation of the temperature dependence of the magnetic susceptibility showed nice 
qualitative agreement with the experimental data.
The field dependence of magnetization, magnetic susceptibility, and entanglement entropy revealed a clear correlation between magnetic and quantum properties. In particular, the correlation between magnetization jumps and susceptibility peaks indicates a strong interplay between magnetic properties and quantum entanglement entropy in both tetranuclear and pentanuclear $Cu^{II}$ complexes. These results indicate that field-induced level crossings and quantum phase transitions play a central role in determining both the thermodynamic properties and the entanglement entropy as functions of the external magnetic field.

Nearest-neighbor entanglement dominates at low temperatures, while next-nearest-neighbor correlations emerge near critical fields. The Dzyaloshinskii–Moriya interaction induces anisotropy, modifying the magnetic response, and enhancing entanglement in specific regimes.

These results can inform future experimental studies of the properties of tetranuclear and pentanuclear copper(II) complexes under stronger magnetic fields with antiferromagnetic interactions.
Further research could focus on the study of antiferromagnetic star-shaped heterogen (heterometallic) clusters.
This study highlights the potential of molecular spin clusters as controllable quantum systems and provides useful insights for the design and interpretation of experiments in molecular magnetism. Moreover, given the possibility of implementing star graphs in solid-state setups such as superconducting networks \cite{Lucci2022} and ultracold atoms in optical lattices \cite{Brunelli2004,Buccheri2016,Burioni2000}, our results may provide a theoretical counterpart for studies of quantum correlations in regimes where the molecules or stars have a small number of sites.

\section*{Acknowledgments}
N. A., G. M., and L. A.  acknowledge the financial support provided within the framework of  project No.~SCS~21AG-1C006. N. T. acknowledges funding from the European Union’s Horizon Europe Research and Innovation Programme under the Marie Sk\l{}odowska-Curie Doctoral Network MAWI (Matter-Wave Interferometers) under the grant agreement No. 101073088.

% Uncomment and use as the case may be
%\begin{theorem} 
%\end{theorem}

% Uncomment and use as the case may be
%\begin{lemma} 
%\end{lemma}


%\section{}\label{}


%\bibliographystyle{cas-model2-names}

%\bibliography{cas-refs}

\end{document}